\documentclass[12pt]{article}
\usepackage{amssymb}
\usepackage{amsfonts}
\usepackage{color}
\usepackage{amsmath,amsthm}
\usepackage{graphics,epsfig}
\usepackage{graphicx,epsfig, color }
\oddsidemargin 10mm \numberwithin{equation}{section}
\def\be{\begin{equation}}
\def\ee{\end{equation}}
\begin{document}
\begin{center} {{\bf {Quintessence Reissner Nordstr\"{o}m Anti de Sitter Black Holes and Joule Thomson effect}}\\
 \vskip 0.50 cm
  {{ H. Ghaffarnejad \footnote{E-mail: hghafarnejad@semnan.ac.ir
 } }{ E. Yaraie \footnote{E-mail: eyaraie@semnan.ac.ir
 } }{ M. Farsam \footnote{E-mail: mhdfarsam@semnan.ac.ir
 } }}\vskip 0.2 cm \textit{Faculty of Physics, Semnan
University, 35131-19111, Semnan, Iran }}
\end{center}
\begin{abstract}
In this work we investigate corrections of the quintessence regime
of the dark energy on the Joule-Thomson (JT) effect of the
Reissner Nordstr\"{o}m anti de Sitter (RNAdS) black hole.  The
quintessence dark energy has equation of state as
$p_q=\omega\rho_q$ in which $-1<\omega<-\frac{1}{3}.$ Our
calculations are restricted to ansatz: $\omega=-1$ ( the
cosmological constant regime ) and $\omega=-\frac{2}{3}$
(quintessence dark energy). To study the JT expansion of the AdS
gas under the constant black hole mass, we calculate inversion
temperature $T_i$ of the quintessence RNAdS black hole where its
cooling phase is changed to heating phase at a particular
(inverse) pressure $P_i.$ Position of the inverse point
$\{T_i,P_i\}$ is determined by crossing the inverse curves with
the corresponding Gibbons-Hawking temperature on the T-P plan. We
determine position of the inverse point verse different numerical
values of the mass $M$ and the charge $Q$ of the quintessence AdS
RN black hole. The cooling-heating phase transition (JT effect) is
happened for $M>Q$ in which the causal singularity is still
covered by the horizon. Our calculations show sensitivity of the
inverse point $\{T_i,P_i\}$ position on the T-P plan to existence
of the quintessence dark energy just for large numerical values of
the AdS RN black holes charge $Q$. In other words the quintessence
dark energy dose not affects on position of the inverse point when
the AdS RN black hole takes on small charges.
\end{abstract}
\section{Introduction}
Astronomical observation shows the accelerating expansion of the
Universe [1-3].  The origin of the acceleration comes from
negative pressure  which can be caused by two different factors:
(a) the cosmological constant, (b) the quintessence hypothetical
form of the dark energy. The latter cosmic source has an equation
of state as $p_q=\omega \rho_q$ in which the subscript $q$ denotes
to the word `quintessence` and $\omega$ is barotropic index such
that $-1<\omega<-\frac{1}{3}$ [4-7]. Origin of the quintessence
can be made from dynamical scalar fields [8,9]. The border state
of the quintessence namely $\omega=-1$  covers the cosmological
constant regime. The cosmological constant affects on horizon of
the black holes. For instance its effect on spherically symmetric
static metric is well known as Schwarzschild-de Sitter black hole
[10]. While the acceleration effects of the Universe caused by the
quintessence is studied in [11,12]. Many authors are studied the
effects of the quintessence dark energy on the black holes. For
instance, Chen et al are studied effects of the quintessence dark
energy on the Hawking radiation and quasi normal modes of black
holes [13,14].
 According to
the original work [15] we know that in the extended phase space,
the cosmological constant treats as thermodynamic pressure
$P=-\frac{\Lambda}{8\pi}$ and its conjugate quantity acts as
thermodynamic volume $V=\big(\frac{\partial M}{\partial
P}\big)_{S,Q}$ where $S,Q,M$ are the entropy, the charge and the
mass of an AdS black hole respectively. AdS background is a vacuum
de Sitter space time where the cosmological constant has negative
values. See also [16] where  Kubiznak and Mann are studied the
phase structure of the quintessence RNAdS black hole in the
extended phase space. In general, thermodynamic volume is
different with a geometrical volume of the black holes but for RN
type they become similar [17]. Black hole mass  $M$ treats also as
enthalpy of the AdS black hole in the extended phase space
[18-21]. Readers can be follow relationship between the extended
phase space with holographic heat engines perspective for charged
AdS black holes in ref. [22]. Charged AdS black holes treat as
liquid-gas system. For instance a RNAdS black hole behaves as the
Van der Waals liquid gas with first order phase transition [23,
24].  Cai et al are studied P-V criticality in the extended phase
space of Gauss-Bonnet black holes in AdS space time [25]. They
obtained  a P-V criticality and the large-small black hole phase
transition just for a 5 dimensional black hole with
 spherical horizon. There is not obtained the P-V
criticality for Ricci flat and hyperbolic Gauss Bonnet black holes
in 5 dimension. Heat engine of a charged AdS black hole related to
a small-large black hole phase transition is also studied in ref.
[26].  See also [27] where the influence of the quintessence dark
energy is studied on the holographic thermalization of the
gravitational collapse. Superconductor formation of a quintessence
RNAdS black hole is studied also in ref. [28] by using the
holographic framework. Two point correlation functions which are
made from bilinear quantum matter field operators, are well known
as non-local observable. Their thermodynamical properties are
obtained to be similar to the entanglement entropy [29-34] in
which the authors showed both of them take on  similar
non-equilibrium thermalization behavior. Authors of the works
[35-42] showed that the entanglement entropy and two point
functions behave as the similar superconductor phase transition
and in refs. [43,44] is shown that they have similar cosmological
singularity. Authors of [45] are studied phase structure of the
quintessence RNAdS black hole by the nonlocal observable such as
holographic entanglement entropy and two point correlation
function. They obtained Van der Waals-like phase transition. In
short, we can use entanglement entropy approach as a good probe to
study the aspects of holographic superconductors and investigate
phase transitions more deeply. Existence of the small-large AdS
black hole phase transition comes from Hawking-page phase
transition [46] where a large AdS black hole is stable while a
small AdS black hole is unstable because of quantum matter
interactions. This phenomena leads us to study cooling-heating
regime of an AdS black hole. Instability of small AdS black holes
reach to a stable gas in AdS background finally. Usually the gas
exhibits with a Joule-Thompson (JT) expansion for which gas at a
high pressure reaches to a low pressure under the constant
enthalpy (the black hole mass). \"{O}kc\"{u} and
 Aydiner, studied the JT expansion for AdS charged black hole and
obtained some similarities and differences with Wan der Waals
fluid [47] and they extended  their work to a Kerr-AdS black
 hole in ref. [48].
 In general the JT expansion of a gas is happened
at constant enthalpy which for a black hole is its mass.
 JT expansion [49] is an actual iso-enthalpic
thermodynamical experiment in which a thermal system exhibits with
a thermal expansion via molecular interaction. The JT coefficient
$\mu_{JT}=\big(\frac{\partial T}{\partial P}\big)_H$ is a suitable
parameter which determines cooling or heating phase of the system.
In a gas expansion with temperature $T$, the pressure decreases,
so the sign of $\partial P$
 is negative by definition. Hence we can
 consider two different state by defining  inverse temperature $T_i$ which is computed from the equation
 $\mu_{JT}(T_i)=0$:
If $T<T_i (T>T_i)$ for which $\partial P<0$ by definition, then
the JT effect cools (warms) the gas with $\partial T<0 (>0)$ and
so $\mu_{JT}>0 (<0)$. When the gas takes on the inverse
temperature $T_i$ it will be have an inverse pressure which we
call here $P_i.$ Cooling-heating process is happened at the
inverse point $(T_i,P_i)$ (see figures 1,2,3) on the T-P phase
space for a gas expansion because of molecular interaction with a
constant enthalpy. For instance we have $\mu_{JT}=0$ for ideal gas
 because of no interaction between the molecules [49].
In the present work  we study quintessence dark energy effects on
the  JT expansion of a RNAdS black hole under the conditions of
the constant mass and the constant charge. In short one can infer
by looking to diagrams of the figures 1,2,3 that output of our
work is: (a) location of the inverse point $(T_i,P_i)$ is obtained
by crossing diagrams of the inverse curve $T_i(P_i)$ and diagrams
of the state equation $T(P).$ Its position is located where the
temperature of the AdS black hole gas takes on its maximum value.
We should notice that the inverse point is a critical point on the
T-P phase space where the quintessence RNAdS black hole heating
phase separates with its cooling phase. (b) The JT expansion is
happened at high (low) pressure-temperature for small (large)
fixed charge and mass parameter in absence and presence of the
quintessence dark energy effect. (c) The quintessence dark energy
effects  become negligible on the JT expansion of the AdS gas for
small numerical values of the black hole charge but should be
considered more for large numerical values of the charge.
\\ Organization of the work is as follows. In Section 2 we
introduce RNAdS black hole metric surrounded with quintessence
dark energy by according to the work presented by Kiselev [50]
(see also [45]). In section 3 we calculate the JT coefficient of
the quintessence RNAdS black hole and determine its cooling and
heating phase. Section 3 denotes to conclusion and outlook.
\section{Quintessence RNAdS Black Holes}
Applying additivity and linearity conditions on the quintessence
dark energy stress tensor one can infer [50]
\begin{equation} T_{t}^{t}=T_{r}^{r}=\rho_{q},
\end{equation}
\begin{equation}
T_{\theta}^{\theta}=T_{\phi}^{\phi}=-\frac{1}{2}\rho_{q}(3\omega+1)
\end{equation}  where $\rho_q(r)=-\frac{a}{2}\frac{3\omega}{r^{3(1+\omega)}}.$ $\omega$ is state parameter and  $a$ is the
normalization factor related to the density of quintessence. For
quintessence (phantom) regime of the dark energy the barotropic
index will have $-1<\omega<-\frac{1}{3} (\omega<-1)$ and so
positivity condition of the energy density $\rho_q$ is satisfied
by choosing $a>0 (a<0).$ Applying (2.1) and (2.2) and units $4\pi
G=1,$ Kiselev solved the Einstein-Maxwell metric equation
$G_{\mu\nu}+\Lambda g_{\mu\nu}=8\pi
G\{T_{\mu\nu}^{(maxwell)}+T_{\mu\nu}^{(quintessence)}\}$ in the
presence of the cosmological constant $\Lambda.$ He obtained
series solutions for metric of the spherically symmetric static
black hole surrounded by the quintessence dark energy (see Eq. 18
in Ref. [50]). Leading order part of his series solution, reduces
to the quintessence RNAdS black hole metric which is given by (see
Eq. 21 in Ref. [50].) \be
ds^2\approx-f(r)dt^2+\frac{dr^2}{f(r)}+r^2(d\theta^2+\sin^2\theta
d\varphi^2)\ee with \be
f(r)=1-\frac{2M}{r}+\frac{Q^2}{r^2}+\frac{r^2}{l^2}-\frac{a}{r^{3\omega+1}}\ee
where $M$ is the black hole mass. $l$ is radius of the AdS space
time which is related to the cosmological constant with negative
value which for $`n`$ dimensional AdS space time become
$\Lambda=-\frac{(n-1)(n-2)}{2l^2}<0$ [27,51]. One can obtain
modified mass $\widetilde{M}=M+\frac{a}{2}$ for ansatz $\omega=0,$
modified charge $\widetilde{Q^2}=Q^2-a$ for ansatz
$\omega=\frac{1}{3}$ and modified AdS radius
$\frac{1}{\widetilde{l^2}}=\frac{1}{l^2}-\frac{1}{a}$ for ansatz
$\omega=-1.$ In general,
 one can infer
 that the free quintessence  generates  the horizon of the black hole if $0<3\omega+1<1$ for which
 \be-\frac{1}{3}<\omega<0
\ee and generates  the AdS radius $l$ if $-2<3\omega+1<-1$ where
\be -1<\omega<-\frac{2}{3}.\ee  Kiselev is also calculated affects
of the quintessence dark energy on the Gibbons-Hawking temperature
of the RNdS black hole in ref. [50] where the quintessence
decreases temperature of the RNdS black hole which vanishes for
$a=(8M)^{3\omega+1}$. Phase structure of the above black hole
solution is investigated by applying the nonlocal observable in
ref. [45]. The authors obtained a Van der Waals phase transition
for the black hole metric (2.4) in the holographic framework. They
check the equal area law for the first order phase transition and
critical exponent of the heat capacity for the second order phase
transition against different values of $\omega.$ They also discuss
the effect of the barotropic parameter $\omega$ on the phase
structure of the nonlocal observables. Now we study JT expansion
of the quintessence RNAdS black hole as follows.
\section{Joule Thompson effect}
As we said in the introduction section the Hawking and the Page
showed in ref. [46] that there is a critical temperature where a
large stable black hole in AdS space time reaches to thermal gas
in the AdS background. This predicts a large-small black hole
phase transition. According to this work some authors took action
to study thermodynamic properties of AdS black holes due to the
AdS/CFT correspondence [52,53,54] in which the black holes
thermodynamics is identified with the dual strongly coupled CFT in
the boundary of the bulk AdS space time. Thermodynamic property of
an AdS black hole is quite different from those of the
asymptotically flat or de Sitter space time. Large AdS black holes
are thermodynamically stable while small ones are unstable.
Instability of the small AdS black holes exhibit the Joule Thomson
expansion of the AdS gas under the condition of constant enthalpy.
The JT expansion of the gas has an important coefficient
$\mu_{JT}$ which determines cooling and/or heating phase of
iso-enthalpic expansion of the gas. In general, the JT coefficient
is called by definition as [49] \be \mu_{JT}=\bigg(\frac{\partial
T_{b}}{\partial
P}\bigg)_H=\frac{V}{C_P}\bigg(\frac{T_{b}}{T_{b}^i}-1\bigg)\ee in
which $C_P=\big(\frac{\partial H}{\partial T_{b}}\big)_P$ is heat
capacity at constant pressure and $T_{b}^i=V\big(\frac{\partial
T_{b}}{\partial V}\big)_{P_i}$ is inversion temperature for which
$V$ is thermodynamic volume of the gas (see introduction).
 One can infer $\mu_{JT}(T^{b}_i)=0$ where there is no molecular interaction of the gas.
  Applying the thermodynamic equation $H=PV+U$ one can rewrite
 the equation (3.1) as
\be\mu_{JT}=\frac{T_{b}-T_{b}^i}{P+\big(\frac{\partial U}{\partial
V}\big)_P}\ee   for which \be C_P=\big(\frac{\partial M}{\partial
T_{b}}\big)_P=\frac{V}{T_{b}^i}\big[P+\big(\frac{\partial
U}{\partial V}\big)_P\big].\ee  As we said in the introduction
section the above JT equation can be applicable for an AdS black
hole because of extension of the phase space. In the latter case
the negative cosmological constant related to the AdS radius $l$
treats as thermodynamical pressure and its conjugate acts as
thermodynamic volume and the black hole mass behaves as the
enthalpy. According to the latter statements the pressure $P$
given in the above equations should correspond to the AdS radius
of the 4 dimensional space time (2.3) as follows [17,19,55].
\begin{equation}
P=\frac{3}{8\pi l^2}.
\end{equation}
The above identity shows that the AdS radius $l$ plays as a
thermal variable. Location of exterior horizon $r_{+}$ of the
black hole metric (2.3) is determined by choosing the largest root
of the equation $f(r_+)=0$ given by (2.4). To do so we can use
(3.4) and write \begin{equation}
M=\frac{r_+}{2}-\frac{a}{2r_+^{3\omega}}+\frac{4\pi
P}{3}r_+^3+\frac{Q^2}{2r_+}
\end{equation}    which by according to the work
[19] means the AdS black hole enthalpy. In the thermodynamic
perspective the equation (3.5) reads
 \be M=H=PV+U\ee where
$V=\big(\frac{\partial M}{\partial P}
 \big)_{S, Q}=\frac{4\pi}{3} r_+^3$ is the thermodynamic volume of AdS RN black hole surrounded
 with the quintessence dark energy (see the introduction section) and \be
U(r_+)=\frac{r_+}{2}+\frac{Q^2}{2r_+}-\frac{a}{2}\frac{1}{r_+^{3\omega}}\ee
is its internal energy. Applying (2.4) and (3.5) one can calculate
Gibbons-Hawking temperature of the quintessence-RNAdS black hole
as
\begin{equation}
T_{b}=\frac{f^{\prime}(r)}{4\pi}\big |_{r_+}=\frac{1}{4\pi
r_{+}^2}\bigg(3a\omega r_{+}^{-3\omega}+8\pi
Pr_{+}^3+r_{+}-\frac{Q^2}{r_+}\bigg).
\end{equation}
   To plot the isoenthalpic curves in
  T-P plane we need an explicit formula for $T(P).$ To do so we should eliminate $r_+$ between (3.5) and (3.8).
  In general, without setting numerical values on the barotropic index $\omega,$
  we can not obtain an analytic solution for $T(P)$, while one
  obtain
   \be -8192P^3\pi^3Q^6-432\pi^4Q^4T^4+9216P^2\pi^2Q^6a-1728M^2P\pi^3Q^2T^2
   $$$$+1152P\pi^3Q^4T^2-
  3456P\pi Q^6a^2-864M^3\pi^3T^3+648M^2\pi^2Q^2T^2a
  $$$$+864M\pi^3Q^2T^3-432\pi^2Q^4T^2a+432Q^6a^3+5184M^4P^2\pi^2
  -6912M^2P^2\pi^2Q^2
  $$$$+1536P^2\pi^2Q^4-3888M^4P\pi a+5184M^2P\pi Q^2a-1152P\pi Q^4a+729M^4a^2$$$$+108M^2\pi^2T^2
  -972M^2Q^2a^2-108\pi^2Q^2T^2+216Q^4a^2$$$$+72M^2P\pi-72P\pi Q^2-27M^2a+27Q^2a=0\ee
  for the  cosmological constant regime of the quintessence
  $\omega=-1$ and
   \be -131072P^3\pi^3Q^6-6912\pi^4Q^4T^4-13824\pi^3Q^4T^3a-27648M^2P\pi^3Q^2T^2$$$$+
   73728MP^2\pi^2Q^4a+18432P\pi^3Q^4T^2-7776\pi^2Q^4T^2a^2-13824M^3\pi^3T^3$$$$+13824M\pi^3Q^2T^3+82944M^4P^2\pi^2-10368M^3\pi^2T^2a-
   110592M^2P^2\pi^2Q^2$$$$+1728M^2P\pi Q^2a^2+12096M\pi^2Q^2T^2a+24576P^2\pi^2Q^4-10368P\pi Q^4a^2$$$$+729Q^4a^4-10368M^3P\pi a+864M^3a^3+
   1728M^2\pi^2T^2+11520MP\pi Q^2a$$$$-972MQ^2a^3-1728\pi^2Q^2T^2+1152M^2P\pi-108M^2a^2-1152P\pi Q^2$$$$+108Q^2a^2=0
      \ee
      for  the quintessence regime of the dark energy $\omega=-\frac{2}{3}.$
 Applying (3.8), one can obtain inversion temperature of the quintessence RNAdS black hole   as follows.  \be T_{b}^i=\frac{r_{+}}{3}
 \bigg(\frac{\partial T_{b}}{\partial r_+}\bigg)_{P=P_i}^{r_+=r_{+i}}=\frac{1}{4\pi
r_{+i}^2}\bigg(-a\omega(3\omega+2)r_{+i}^{-3\omega}+\frac{8\pi
P_i}{3}r_{+i}^3-\frac{r_{+i}}{3}+\frac{Q^2}{r_{+i}}\bigg).\ee When
the Gibbons-Hawking temperature (3.8) reaches to (3.11)  as
$T_{b}=T^i_{b}$ for a constant inversion pressure $P=P_i$ and
particular radius $r_+=r_{+i}$, we will have
\begin{equation} T_{b}^i=\frac{1}{4\pi r_{+i}^2}\bigg(3a\omega
r_{+i}^{-3\omega}+8\pi
P_ir_{+i}^3+r_{+i}-\frac{Q^2}{r_{+i}}\bigg).
\end{equation} Subtracting (3.11) and (3.12) we obtain the following
constraint condition. \be
\frac{a\omega(3\omega+5)}{4\pi}r_{+i}^{-(2+3\omega)}+\frac{4}{3}P_ir_{+i}+\frac{r_{+i}^{-1}}{3\pi}-\frac{Q^2}{2\pi}r_{+i}^{-3}=0.\ee
To plot inversion curves in T-P plane we must be eliminate
$r_{+i}$ between (3.12) and (3.13). In general, without setting
numerical values on the barotropic index $\omega,$
  we can not obtain an analytic solution for $T_i(P_i)$, but we will have \be
-8192P_i^3\pi^3Q^4+6912\pi^4Q^2T_i^4+9216P_i^2\pi^2Q^4a-3456P_i\pi
Q^4a^2+432Q^4a^3$$$$-512P_i^2\pi^2Q^2+384P_i\pi Q^2a
-32\pi^2T_i^2-72Q^2a^2-8P_i\pi+3a=0 \ee for  $\omega=-1$ and \be
-8192P_i^3\pi^3Q^4+6912\pi^4Q^2T_i^4+3456\pi^3Q^2T_i^3a+1152P_i\pi^2Q^2T_ia$$$$-216\pi
Q^2T_ia^3-512P_i^2\pi^2Q^2 +288P_i\pi
Q^2a^2-27Q^2a^4$$$$-32\pi^2T_i^2+4\pi T_ia-8P_i\pi+a^2=0 \ee for
 $\omega=-\frac{2}{3}.$ If we set
$a=0$ then the quintessence dark energy correction is removed from
the RNAdS black hole and so our work reaches to results of the
paper [47] (see figure 1).
  To study quintessence dark energy effects on the JT expansion of the
 RNAdS black hole we should choose $a\neq0.$ Hence we use two different ansatz $a=\frac{1}{4}$ for weak coupling
 quintessence (see figure 2) and $a=10$ for strong coupling quintessence
 (see figure 3). To plot diagrams of the inversion curves and the iso-enthalpic curves we should choose some numerical values
 for the mass $M$ and the charge $Q$ of the quintessence RNAdS black hole.
 We are free to choose numerical values of $M$ and $Q.$
 According to the work [47] we avoid particular hypersurfaces
 in T-P plane which exhibits with
 naked singularity. This leads to the condition $M>Q$ (see also [56]).
 To compare our results with which ones are given in ref.
 [47] for RNAdS black hole in absence of the quintessence dark energy, we use
 $M=1.5,2,2.5,3$ for $Q=1;$ $M=2.5,3,3.5,4$ for $Q=2$; $M=13,14,15,16$ for $Q=10$ and $M=22,23,24,25$ for
 $Q=20.$  Then we plot diagrams of inversion curves ${(P_i,T_i)}$ given by (3.14) and (3.15)
 in figures 1, 2 and 3 (see solid lines) for
 $a=0,$ $a=\frac{1}{4}$ and $a=10$ respectively.
 We plot
 isoenthalpic curves ${(P,T)}$ given by the equations (3.9) and (3.10) in
 figures
1, 2 and 3 (see discontinued lines) by regarding the above
mentioned numerical
  values for $a=0,$ $a=\frac{1}{4}$ and $a=10$ respectively .
  According to definition of the JT
  coefficient (3.1), the black holes always cool (warm) above (below) the inversion curves
  (solid line in the figures 1,2,3) during
  the JT expansion. Diagrams in the figure 1 show cooling and or warming phase of a AdS RN black hole without the quintessence effect
  with arbitrary value for $\omega$, the JT effect is happened for all states where $M>Q.$
  Diagrams of the figure 2 (3) show that for weak (strong) quintessence effect $a=\frac{1}{4}(10)$, the JT effect is happened when
  $\omega=-1,-\frac{2}{3}$ for situations where $M>Q.$ Comparing
  the figures 1,2,3 we can infer that physical effects of the
  quintessence changes position of cooling-heating critical point in TP plane
  to the center $(T,P)\to(0,0)$ just for large values of the black
  hole charge. In other words position of the cooling-heating critical point of a quintessence  AdS RN black hole
  is sensitive more to the black hole charge.
\section{Conclusion and outlook}
In this work we considered quintessence dark energy effects on the
JT expansion of the RNAdS black hole and determine regions where
the black hole takes on the cooling-heating phase under the
condition of constant mass. Our mathematical calculations predict
that the quintessence dark energy affects more to large black
holes $M>Q>>1$ where critical inversion point $(T_i,P_i)$ reaches
to some smaller values.  In this paper we assumed that the
quintessence dark energy originates from classical dynamical
scalar fields. In this case  the black hole thermal
entropy satisfies the Bekenesten-Hawking theorem (entropy is equivalent to the horizon surface area).\\
As a future work we will extend the present work for the quantum
conformal field theories of the AdS black holes. To do so we
consider the JT expansion of the quantum RNAdS black holes
surrounded with the quantum quintessence dark energy. In the
latter case the quintessence dark energy originates from quantum
scalar fields for which we should calculate its two point
correlation function counterpart. It is called as non-local
observable. Its dynamical effects appear in the conformal  anomaly
[57](see also [58,59]). Physical effects of the conformal anomaly
leads to an additional logarithmic term for the black hole entropy
by regarding an high-energy cutoff scale. On the other side this
anomaly affects on the classical black hole metric and perturbs
it. It causes to evaporate the black hole mass (see for instance
[56,60,61,62] and references therein) but its final state reaches
to remnant stable mini black hole. Hence its JT expansion can be
challenging issue. One of applicable ways to do the work is to use
4 dimensional Lovelock gravity [63]
 and or Gauss-Bonnet higher order derivative metric action
(see Also [25,34] and references therein).
  \vskip .5cm
  {\bf References}
\begin{description}
\item[1.] S. Perlmutter et al, ``Supernova Cosmology Project
Collaboration``, Astrophys. J. 517, 565, (1999);
arXiv:astro-ph/9812133
\item[2.] A. G. Riess et al, ``Supernova Search Team
Collaboration, Cosmological Constant,`` Astron. J. 116, 1009
(1998); arXiv:atro-ph/9805201
\item[3.] P. M. Garbavich et al, ``Supernova Limits on the Cosmic Equation of State``, Astrophys. J. 509, 74 (1998);
arXiv:astro-ph/9806396
\item[4.] T. Chiba, ``Quintessence, the Gravitational Constant, and Gravity``,  Phys. Rev. D60, 083508 (1998);
arXiv:astro-ph/9903094
\item[5.] N. A. Bahcall, J. P. Ostriker, S. Perlmutter and P. J.
Steinhardt, ``The Cosmic Triangle: Revealing the State of the
Universe``,  Science 284, 1481 (1999); arXiv:astro-ph/9906463
\item[6.] P. J. Steinhardt, L. M. Wang and I. Zlatev, ``Cosmological Tracking Solutions``,  Phys. Rev.
D59, 123504 (1999); arXiv:astro-ph/9812313
\item[7.] L. M. Wang, R.R. Caldwell, J. P. Ostriker and P. J.
Steinhardt, ``Cosmic Concordance and Quintessence``,  Astrophys.
J. 530, 17 (2000); arXiv:astro-ph/9901388
\item[8.] P. Ratra, and L. Peebles, ``Cosmological consequences of a rolling homogeneous scalar field``,  Phys. Rev. D37, 3406 (1988)
\item[9.] R.R. Caldwell, R. Dave, and P. J. Steinhardt, ``Cosmological Imprint of an Energy Component with General Equation-of-State``,
 Phys. Rev. Lett. 80 1582 (1998)
\item[10.] G. W. Gibbons and S. W. Hawking, ``Cosmological event
horizon, thermodynamics, and particle creation``, Phys. Rev. D15,
2738 (1977)
\item[11.] S. Hellerman, N. Kaloper and L. Susskind, ``String theory and quintessence`` JHEP 0106,
003 (2001); arXiv:hep-th/0104180
\item[12.] W. Fischler, A. Kashani-poor, R. McNees and S. Paban,
``The Acceleration of the Universe, a Challenge for String
Theory``,  JHEP 0107, 003 (2001); arXiv:hep-th/0104181
\item[13.] S. Chen, B. Wang and R. Su, ``Hawking radiation in a
D-dimensional static spherically symmetric black hole surrounded
by quintessence``, Phys. Rev. D77, 124011 (2008)
\item[14.] S. Chen and J. Jing, ``Quasi-normal modes of a black
hole surrounded by quintessence``, Class. Quant. Grav. 22, 4651
(2005)
\item[15.] G. Q. Li, ``Effects of dark energy on P-V criticality
of charged AdS black holes``, Phys. Lett.B735, 256 (2014); arXiv:
gr-qc/1407.0011
\item[16.] D.
Kubiznak and R. B. Mann,`P-V criticality of charged AdS black
holes`, JHEP 1207, 033 (2012); hep-th/1205.0559
\item[17.] B.P. Dolan,`Black holes
and Boyle's law - the thermodynamics of the cosmological
constant`,Mod. Phys. Lett. A, 30, 1540002 (2015)
arxiv:gr-qc/1408.4023.
\item[18.] D. Kastor, S. Ray and J. Traschen, ``Enthalpy and the Mechanics of AdS black holes``, Class. Quant. Grav. 26, 195011
(2009); hep-th/0904.2765
\item[19.] B.P. Dolan, `Where is
the PdV term in the first law of black hole thermodynamics $?$`,
gr-qc/209.1272
\item[20.] B. P. Dolan, ``Pressure and volume in the first law of black hole
thermodynamics``, Class. Quant. Grav. 28, 235017
(2011);gr-qc/1106.6260
\item[21.]E.Caceres, P. H. Nguyen
and J. F. Pedraza, ` Holographic entanglement chemistry`, Phys.
Rev. D 95, 106015 (2017) hep-th/1605.00595
\item[22.] Clifford V.
Johnson, `Holographic Heat Engines`, Class. Quantum. Grav. 31,
205002 (2014)hep-th/1404.5982
\item[23.] A. Chamblin, R.
Emparan, C. V. Johnson and R. C. Myers, `Charged AdS black holes
and Catastrophic holography`, Phys. Rev. D60, 064018 (1999)
\item[24.] A. Chamblin, R.
Emparan, C. V. Johnson and R. C. Myers, `Holography thermodynamics
and fluctuations of charged AdS black holes`, Phys. Rev. D60,
104026 (1999)
\item[25.] R. G. Cai, L.M. Cao, L. Li, R.Q. Yang,
`P-V criticality in the extended phase space of Gauss-Bonnet black
holes in AdS space`, JEHP09, 005 (2013); gr-qc/1303.6233
\item[26.] S.W. Wei and Y.X.
Liu,`Charged AdS black hole heat engines` gr-qc/1708.08176
\item[27.] X. X. Zeng, D. Y. Chen and L. F. Li, `` Holographic
thermalization and gravitational collapse in the space time
dominated by quintessence dark energy,`` Phys. Rev. D91, 046005
(2015); arXiv:hep-th/1408.6632
\item[28.] S. Chen, Q. Pan and J. Jing, ``Holographic
superconductors in quintessence AdS black hole spacetime,`` Class.
Quant. Grav. 30, 145001 (2013); arXiv: gr-qc/1206.2069
\item[29.] V. Balasubramanian et al, `` Thermalization of Strongly
Coupled Field Theories``, Phys. Rev. Lett.106, 191601 (2011);
arXiv:hep-th/1012.4753
\item[30.] V. Balasubramanian et al, `` Holographic Thermalization``, Phys. Rev. D84, 026010 (2011);
arXiv:hep-th/1103.2683
\item[31.] D. Galante and M. Schvellinger, `` Thermalization with
a chemical potential from AdS space,`` JHEP 1207, 096 (2012);
arXiv:hep-th/1205.1548
\item[32.] E. Caeres and A. Kundu, `` Holographic Thermalization
with Chemical Potential,`` JHEP 1209, 055 (2012); hep-th/1205.2354
\item[33.] X. X. Zeng, and B. W. Liu, `` Holographic
Thermalization in Gauss Bonnet gravity, Phys. Lett. B726, 481
(2013); hep-th/ 1305.4841
\item[34.]X. X. Zeng, X. M. Liu and B. W. Liu, ``, Holographic
Thermalization with a chemical potential in Gauss Bonnet gravity,
JHEP 03, 031 (2014); hep-th/1311.0718
\item[35.] T. Albash and C. V. Johnson, `` Holographic Studies of
Entanglement Entropy in Superconductors, JHEP 1205, 079 (2012);
hep-th/1205.2605
\item[36.] R. G. Cai, S. He, L. Li and Y. L. Zhang, ``Holographic
Entanglement Entropy in Insulator/Superconductor Transition``,
JHEP 1207, 088 (2012); hep-th/1203.6620
\item[37.] R. G. Cai, L. Li, L. F. Li and R. K. Su, ``
Entanglement Entropy in Holographic P-Wave
Superconductor/Insulator Model,`` JHEP 1306, 063 (2013);
hep-th/1303-4828
\item[38.] L. F. Li, R. G. Cai, L. Li, and C. Shen, Entanglement
entropy in a Holographic P-Wave superconductor Model,`` Nucl.
Phys. B 894, 15 (2015); hep-th/1310.6239
\item[39.] R. G. Cai, S. He, L. Li and Y. L. Zhang,`` Holographic
Entanglement Entropy in Insulator/Superconductor Transition, JHEP
1207, 088 (2012); hep-th/1203.6620
\item[40.] X. Bai, B. H. Lee, L. Li, J. R. Sun and H. Q. Zhang, ``
Time Evolution of Entanglement Entropy in Quenched Holographic
Superconductors,`` JHEP 040, 66 (2015); hep-th/1412.5500
\item[41.] R. G. Cai, L. Li, L. F. Li and R. Q. Yang,``
Introduction to Holographic Superconductor Models, Sci China-Phys
Mech Astron, 58, 060401 (2015); hep-th/1502.00437
\item[42.] Y. Ling, P. Liu, C. Niu, J. P. Wu and Z. Y. Xian, ``
Holographic Entanglement Entropy Close to Quantum Phase
Transition``, hep-th/1502.03661
\item[43.] N. Engelhardt, T. Hertog and G. T. Horowitz, ``
Holographic Signatures of Cosmological Singularities``, Phys. Rev.
 Lett. 113, 121602 (2014); hep-th/1402.2309
\item[44.] N. Engelhardt, T. Hertog and G. T. Horowitz, `` Further
Holographic Investigations of Big Bang Singularities,`` JHEP 1507,
044 (2015); hep-th/1503.08838  
\item[45.]  X. X. Zeng and L. F. Li, ``Van der Waals phase transition in the framework of holography``,
Phys. Lett. B764, 100 (2017);hep-th/1512.08855
\item[46.] S. W. Hawking and D. N.
Page, `Thermodynamics of black holes in anti-de Sitter space`,
Commun. Math. Phys. 87, 577 (1983)
\item[47.] \"{O}. \"{O}kc\"{u} and E. Aydiner, ``  Joule Thomson Expansion of Charged AdS Black
holes``, Eur. Phys. J. C. 77, 24 (2017), gr-qc/1611.06327
\item[48.]  \"{O}. \"{O}kc\"{u}
and E.
 Aydiner, `Joule-Thompson expansion of Kerr-AdS black holes` arxiv:gr-qc/1709.06426
\item[49.] P. J. Gans,`` Joule Thomson Expansion``, Phys.
Chemistry I, 25, 0651, (1992)
\item[50.]  V.V. Kiselev, Quintessence and black holes, Class. Quant. Grav. 20 1187 (2003).
\item[51.] S. Das, P. Majumdar and R. K. Bhaduri, `` General
logarithmic corrections to black hole entropy`` Class. Quant.
Grav.19, 2355,(2002); hep-th/0111001
\item[52.] J. M. Maldacena,` The Large N Limit of
Superconformal field theories and
supergravity`,Adv.Theor.Math.Phys.2:231-252,1998 hep-th/9711200v3
\item[53.] S.S. Gubser, I.R. Klebanov and A.M.
Polyakov, `Gauge Theory Correlators from Non-Critical String
Theory`,Phys.Lett.B428:105-114,1998 hep-th/9802109
\item[54.] E. Witten, `Anti De Sitter Space And
Holography`, Adv.Theor.Math.Phys.2:253-291,1998hep-th/9802150
\item[55.] B. P. Dolan,`` The cosmological constant and the black hole equation of
state``, Class. Quant. Grav.28, 125020 (2011); gr-qc/1008.5023
 \item[56.] H.
Ghaffarnejad, H. Neyad, M. A. Mojahedi, `` Evaporating quantum
Lukewarm black holes final state from back-reaction corrections of
quantum scalar fields,`` Astrophys. Space Sci. 346, 497 (2013)
\item[57.] C. Holzhey, F. Larsen and F. Wilczek, ``Geometric and Renormalized Entropy in Conformal Field
Theory``, Nucl. Phys. B424, 443, (1994); hep-th/9403108
\item[58.] S. Ryu and T. Takayanagi, `Holographic derivation of entanglement entropy from
AdS/CFT,`` Phys. Rev. Lett. 96, 181602 (2006); hep-th/0603001
\item[59.] S. Ryu and T. Takayanagi, `` Aspects of Holographic
Entanglement Entropy,`` JHEP 0608, 045 (2006); hep-th/0605073
\item[60.] H. Ghaffarnejad, `Quantum field backreaction corrections and remnant stable evaporating Schwarzschild-de Sitter dynamical black hole``, Phys. Rev. D75, 084009 (2007)
\item[61.] H. Ghafarnejad,`` Stability of the evaporating Schwarzschild-de Sitter black hole final
state``, Phys. Rev. D74, 104012 (2006)
\item[62.] H. Ghaffarnejad and H. Salehi, `` Hadamard renormalization, conformal anomaly and cosmological event
horizons``, Phys. Rev. D56, 4633 (1997); 57, 5311(E) (1998).
\item[63.] D. Lovelock, `The Einstein tensor and its generalizations`, J. Math. Phys. 12, 3, 498 (1971).
\end{description}
\begin{figure}[tbp] \centering
 \includegraphics[width=6cm,height=4.5cm]{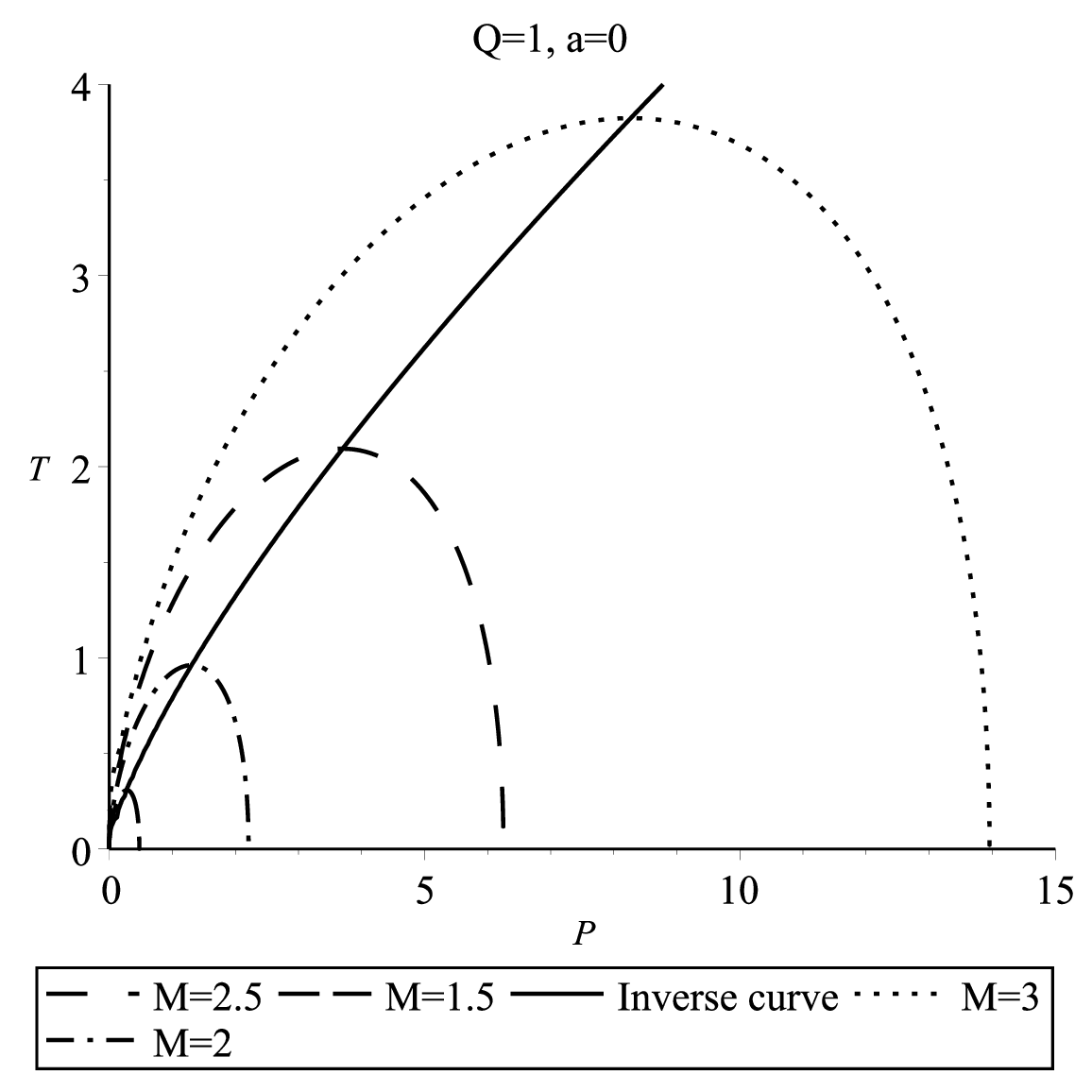}
 \includegraphics[width=6cm,height=4.5cm]{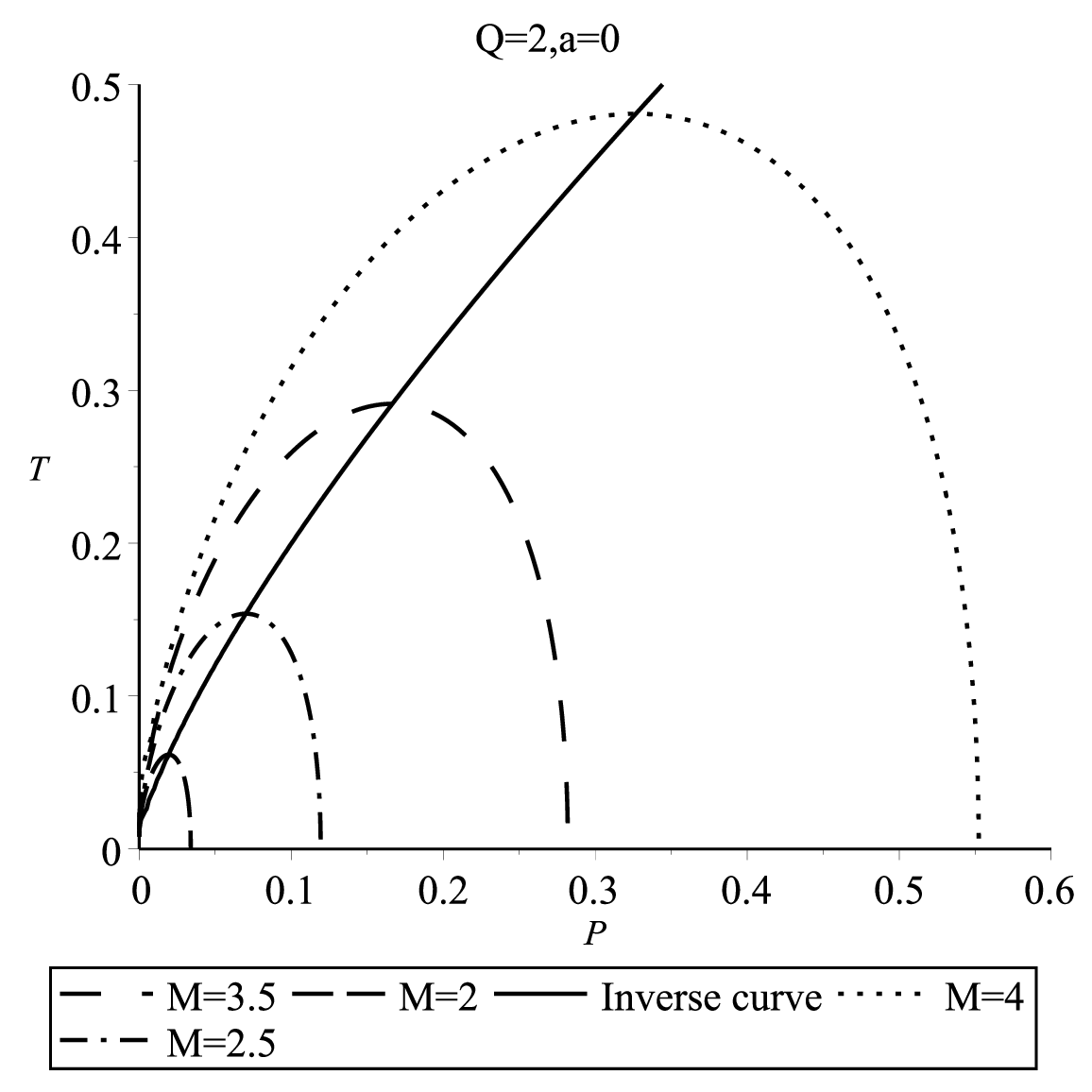}
 \includegraphics[width=6cm,height=4.5cm]{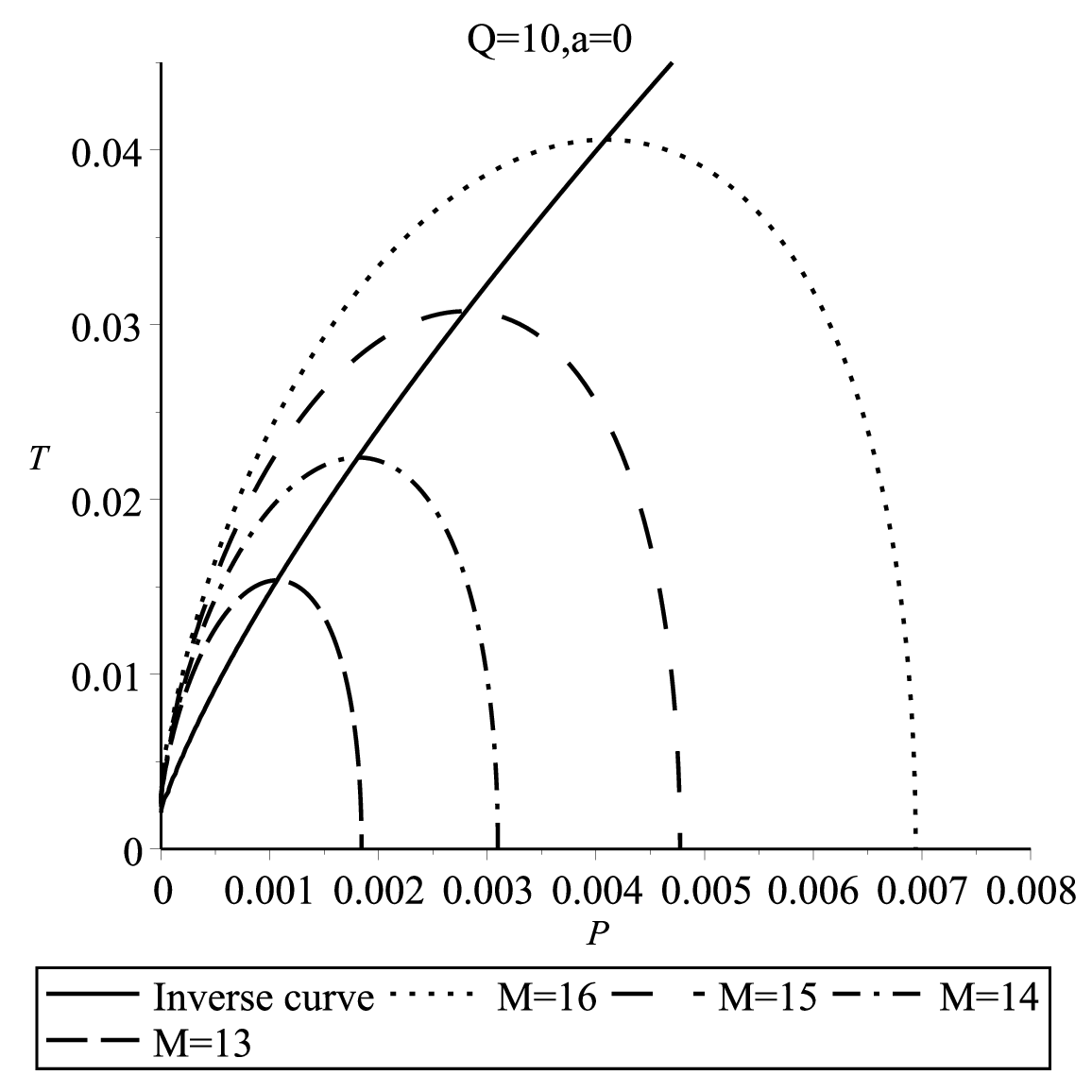}
 \includegraphics[width=6cm,height=4.5cm]{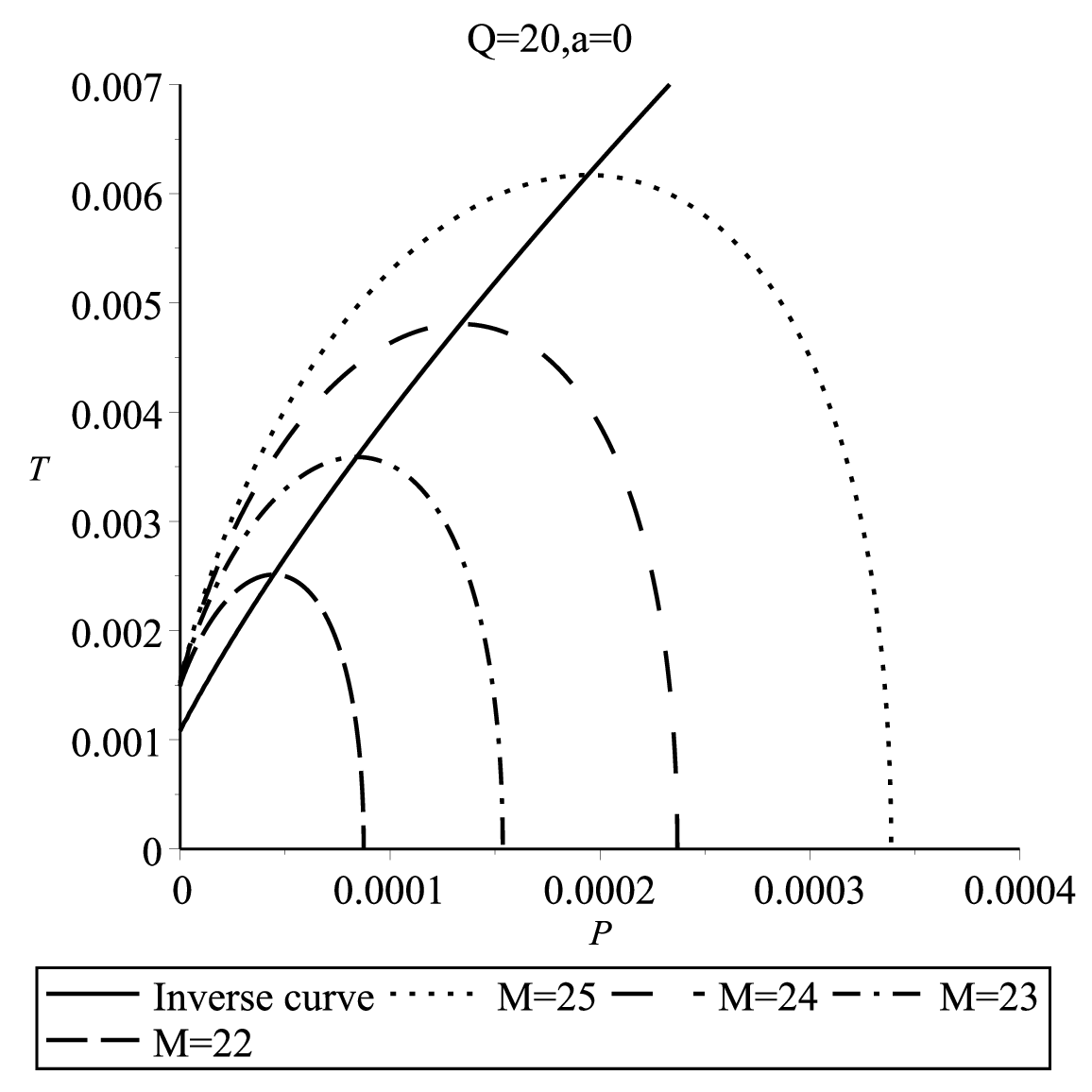}
\caption{\label{fig:1}   Diagrams of the inversion curves and the
isoenthalpic curves  of the RNAdS black hole without  the
quintessence effects $a=0.$}
\end{figure}
\begin{figure}[tbp] \centering
 \includegraphics[width=6cm,height=4.5cm]{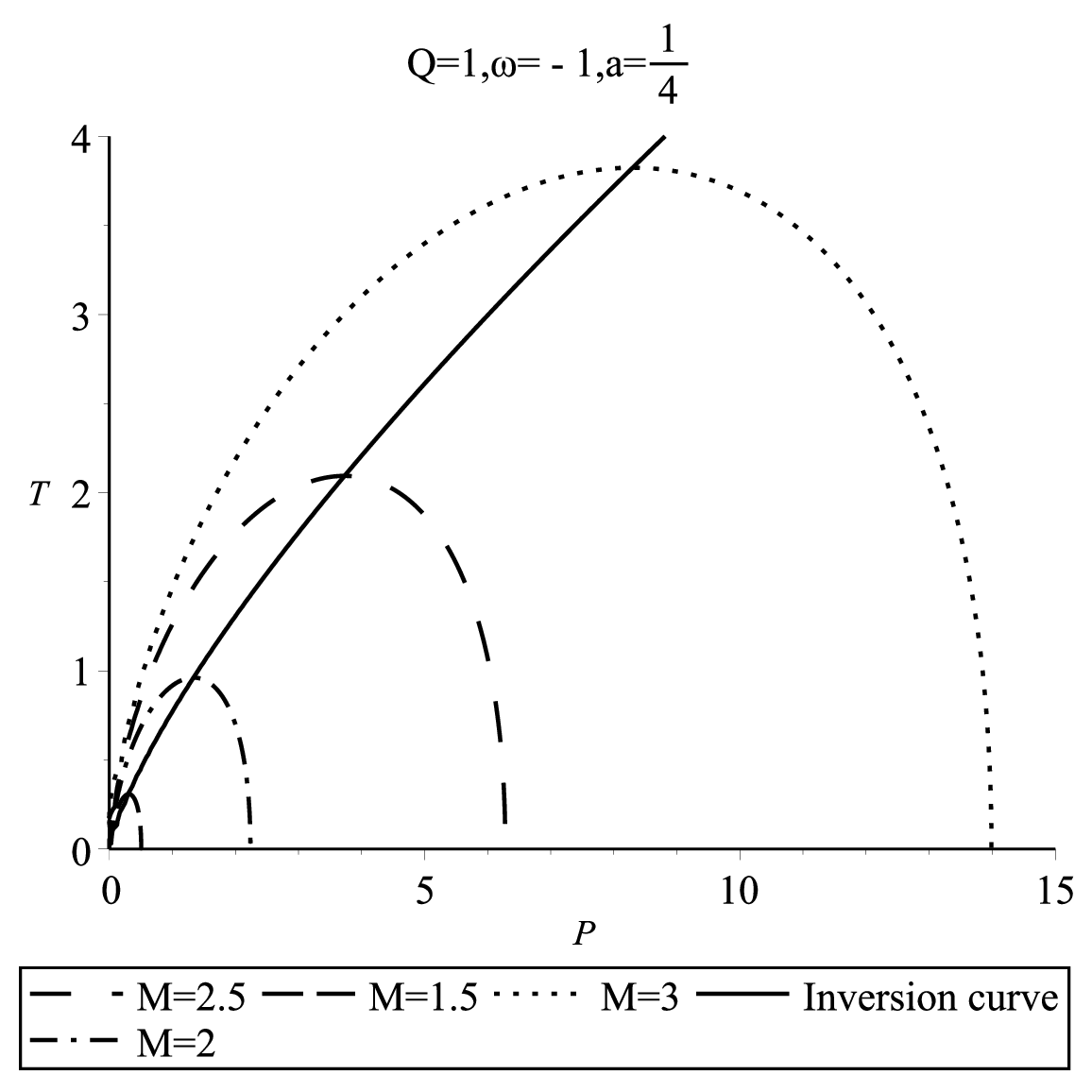}
 \includegraphics[width=6cm,height=4.5cm]{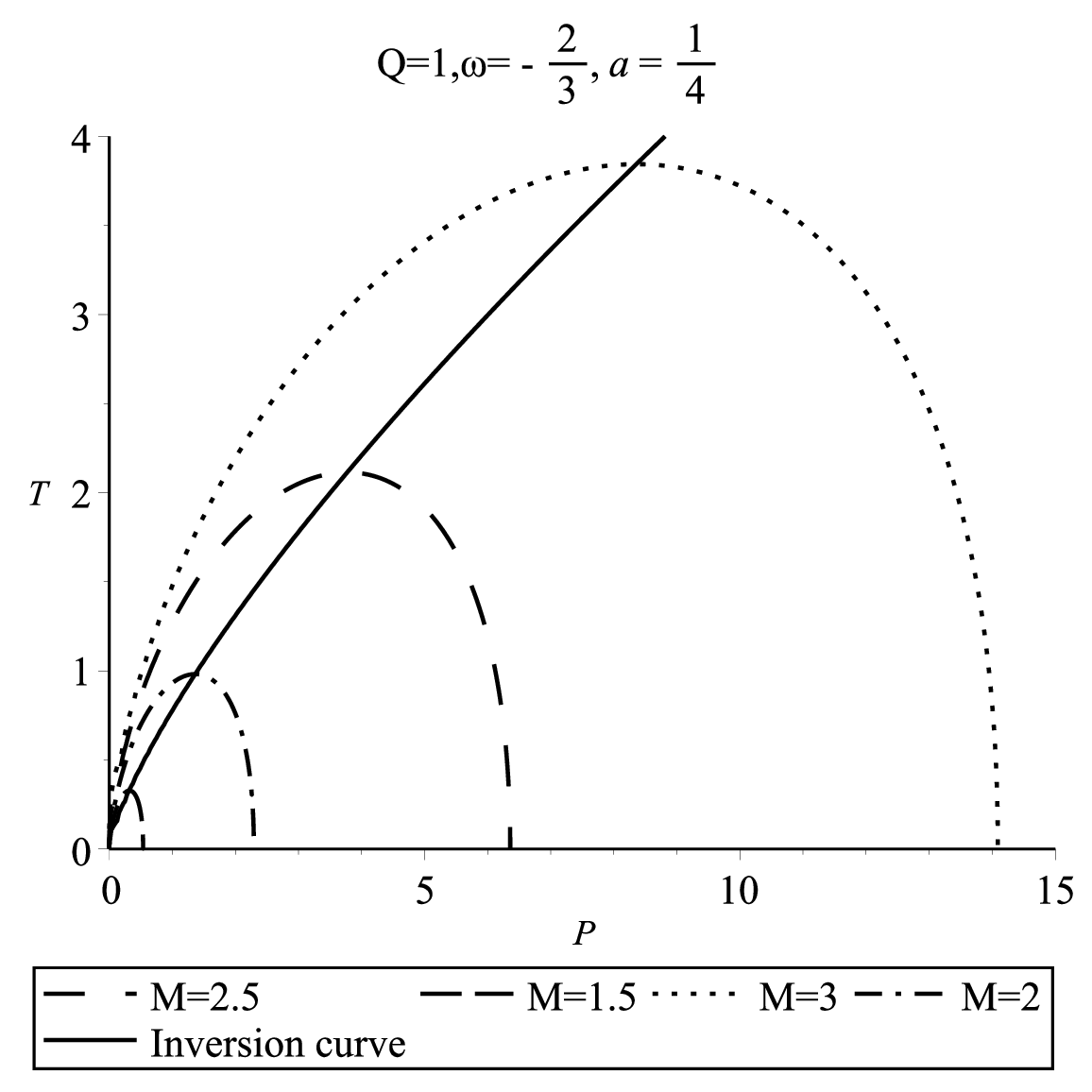}
 \includegraphics[width=6cm,height=4.5cm]{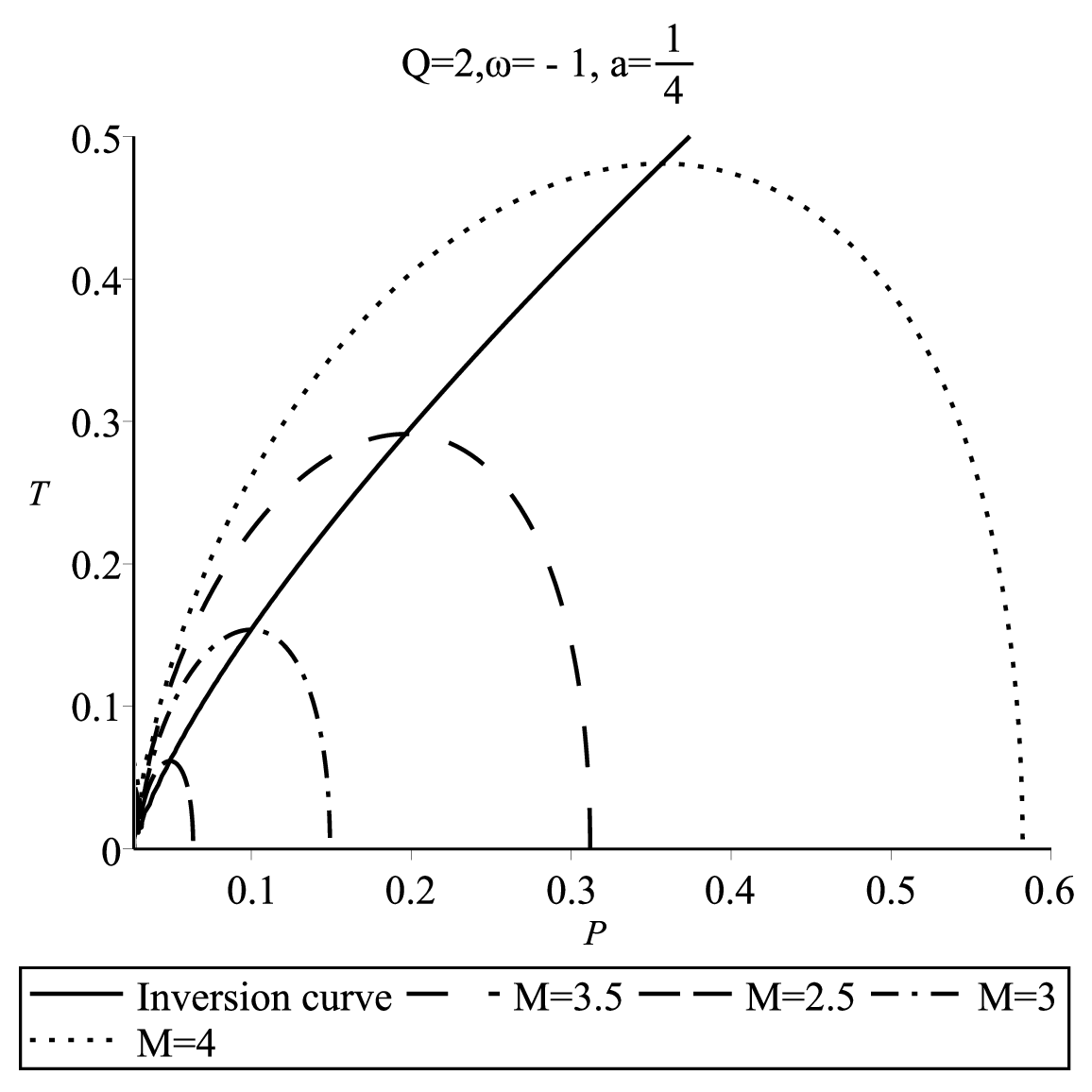}
 \includegraphics[width=6cm,height=4.5cm]{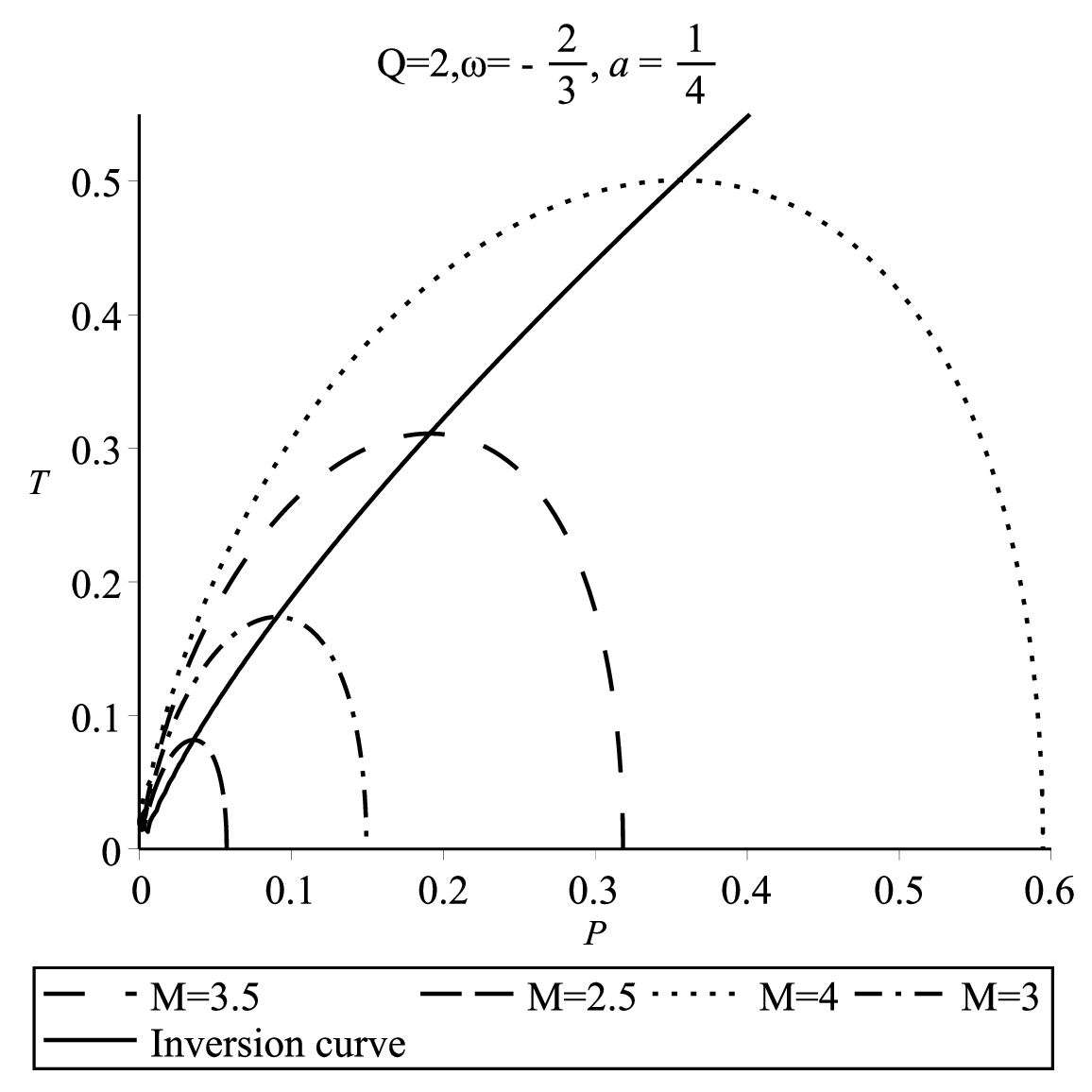}
 \includegraphics[width=6cm,height=4.5cm]{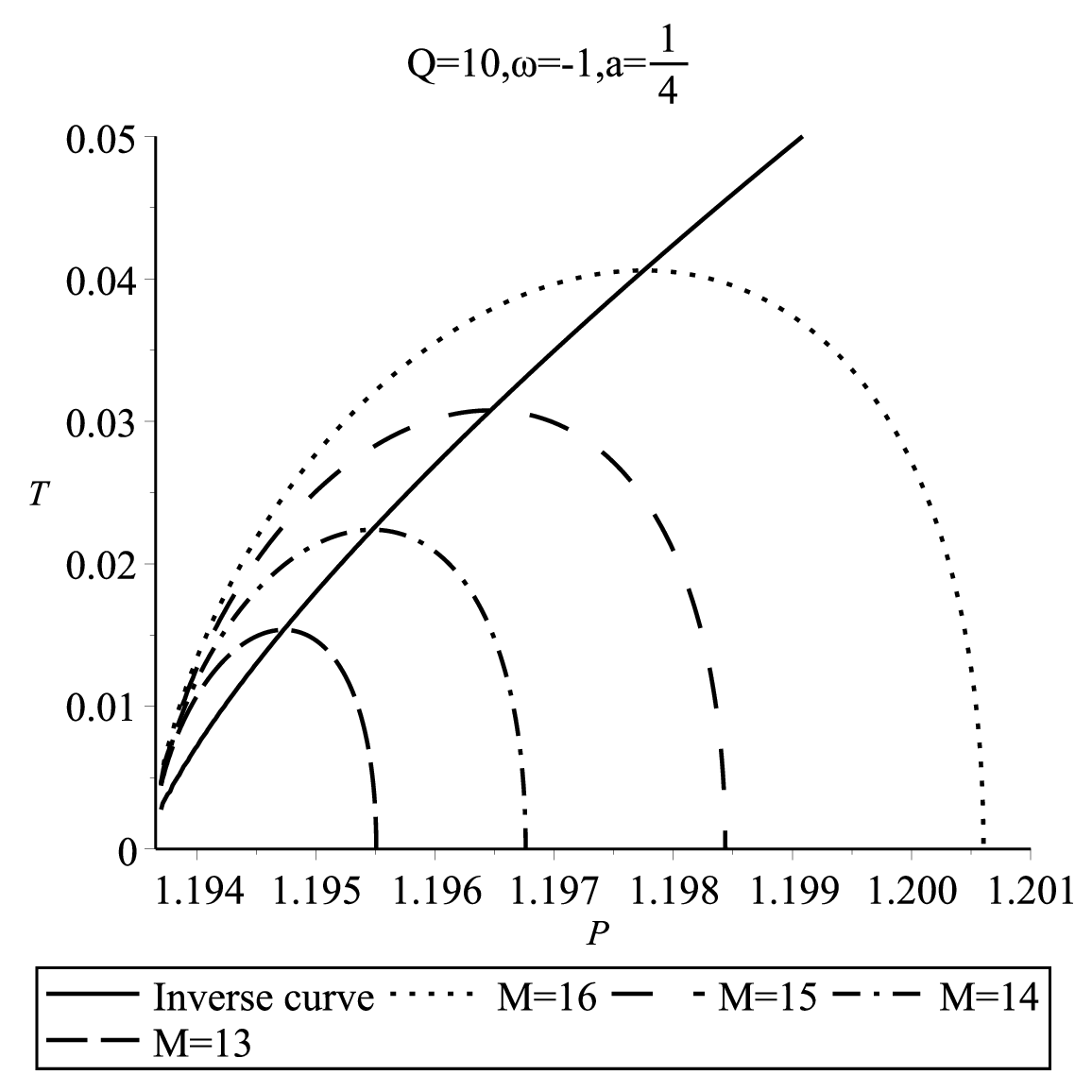}
\includegraphics[width=6cm,height=4.5cm]{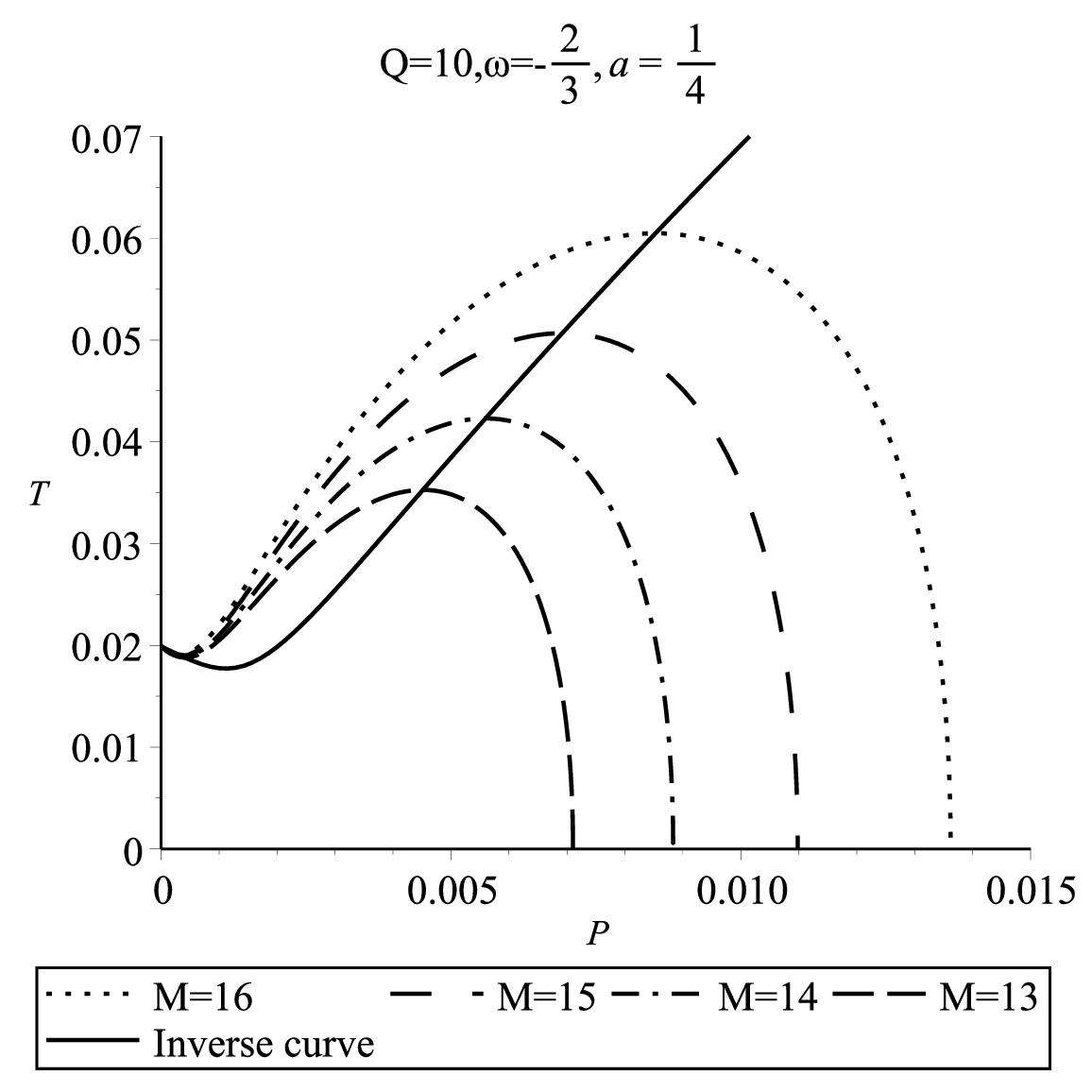}
 \includegraphics[width=6cm,height=4.5cm]{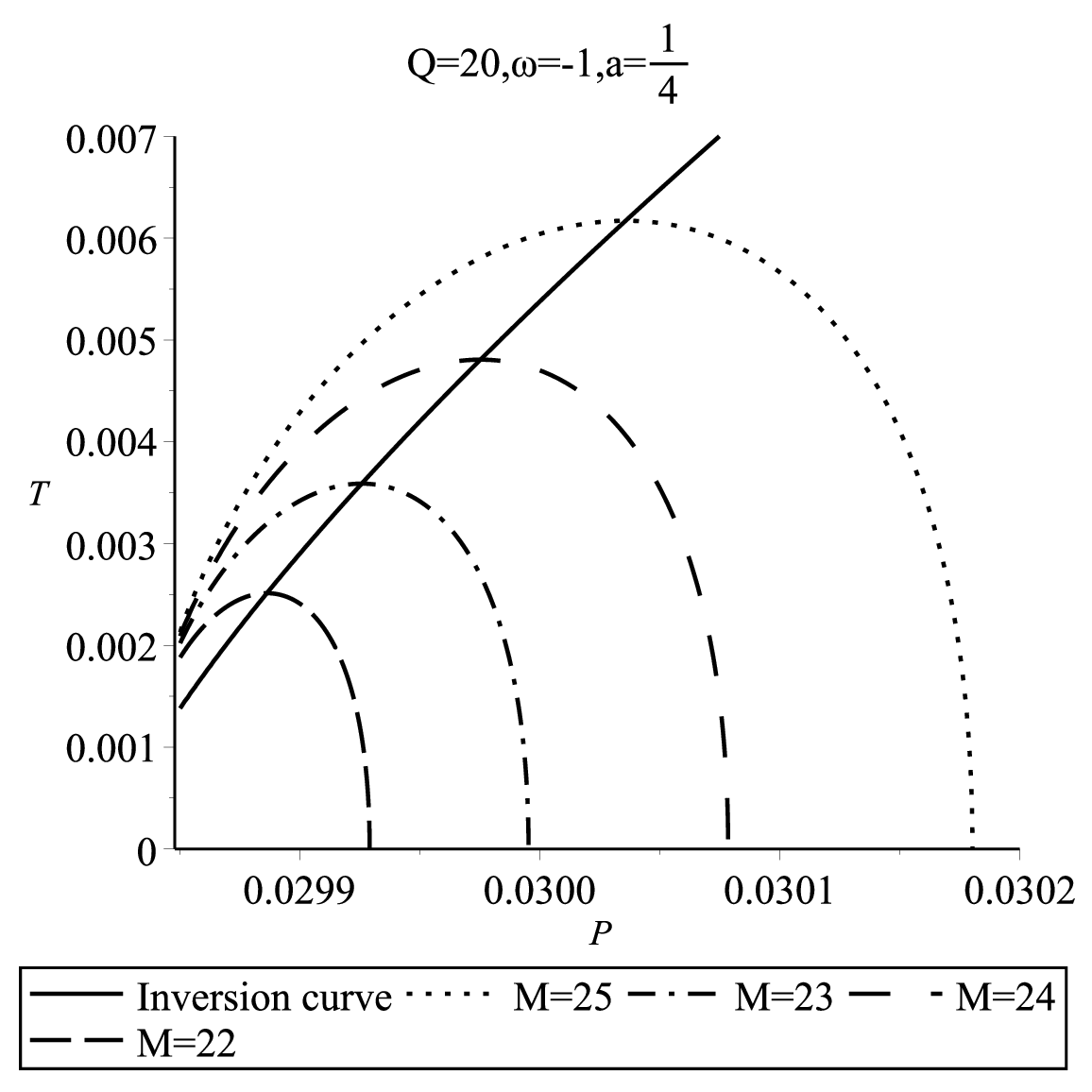}
\includegraphics[width=6cm,height=4.5cm]{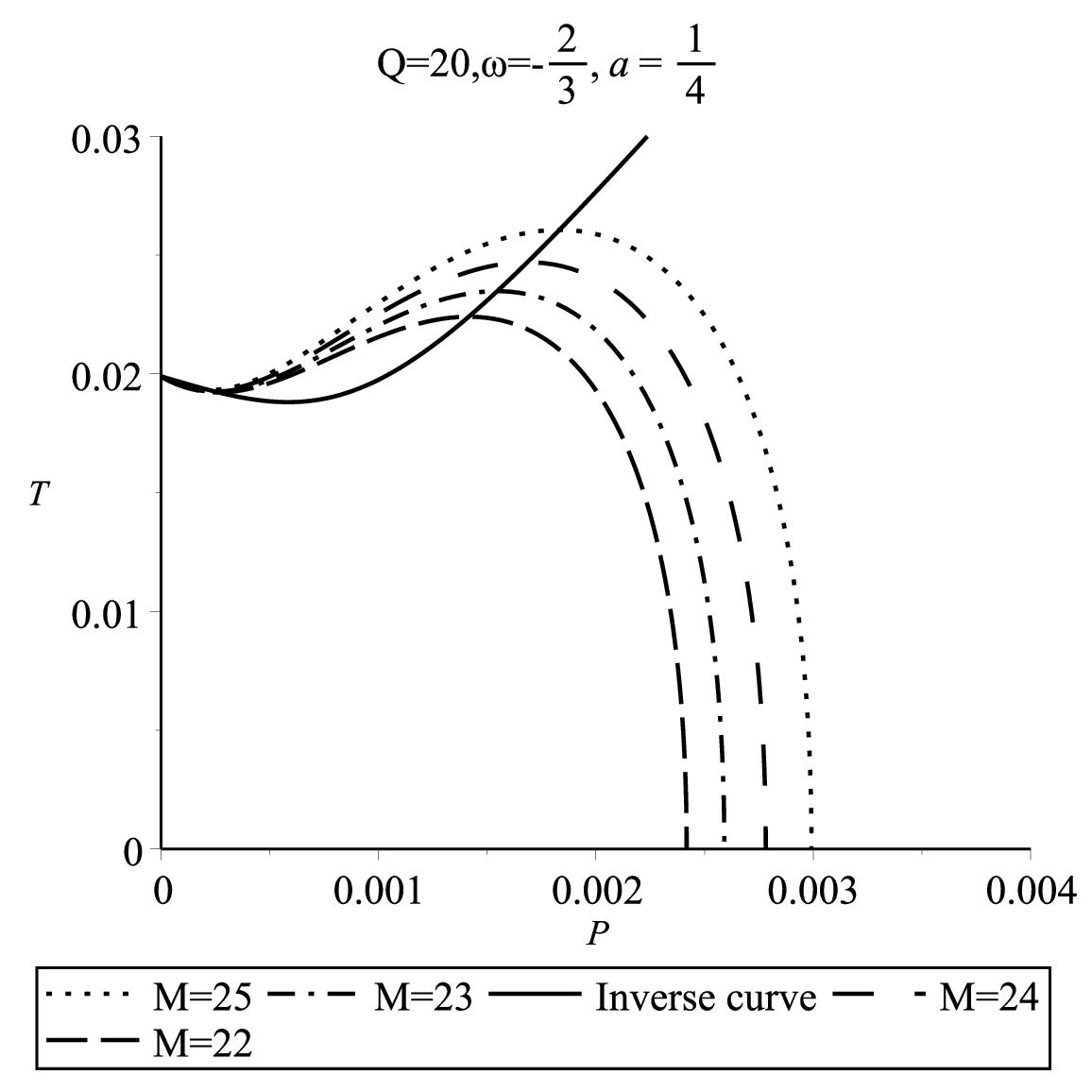}
\caption{\label{fig:1}  Diagrams of the inversion curves and the
isoenthalpic curves  of the  RNAdS black hole with the weak
 effects of the quintessence $a=\frac{1}{4}.$}
\end{figure}
\begin{figure}[tbp] \centering
 \includegraphics[width=6cm,height=4.5cm]{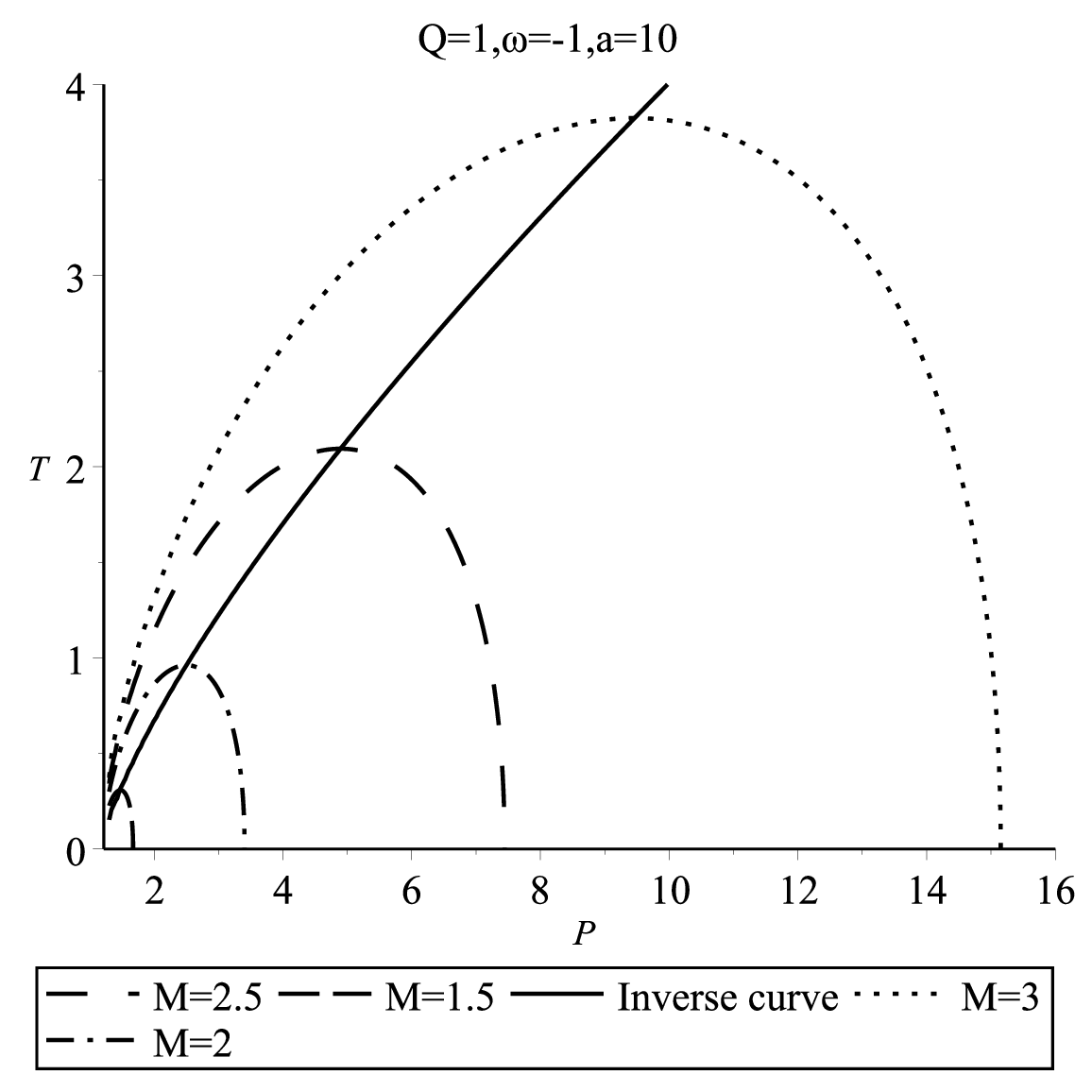}
 \includegraphics[width=6cm,height=4.5cm]{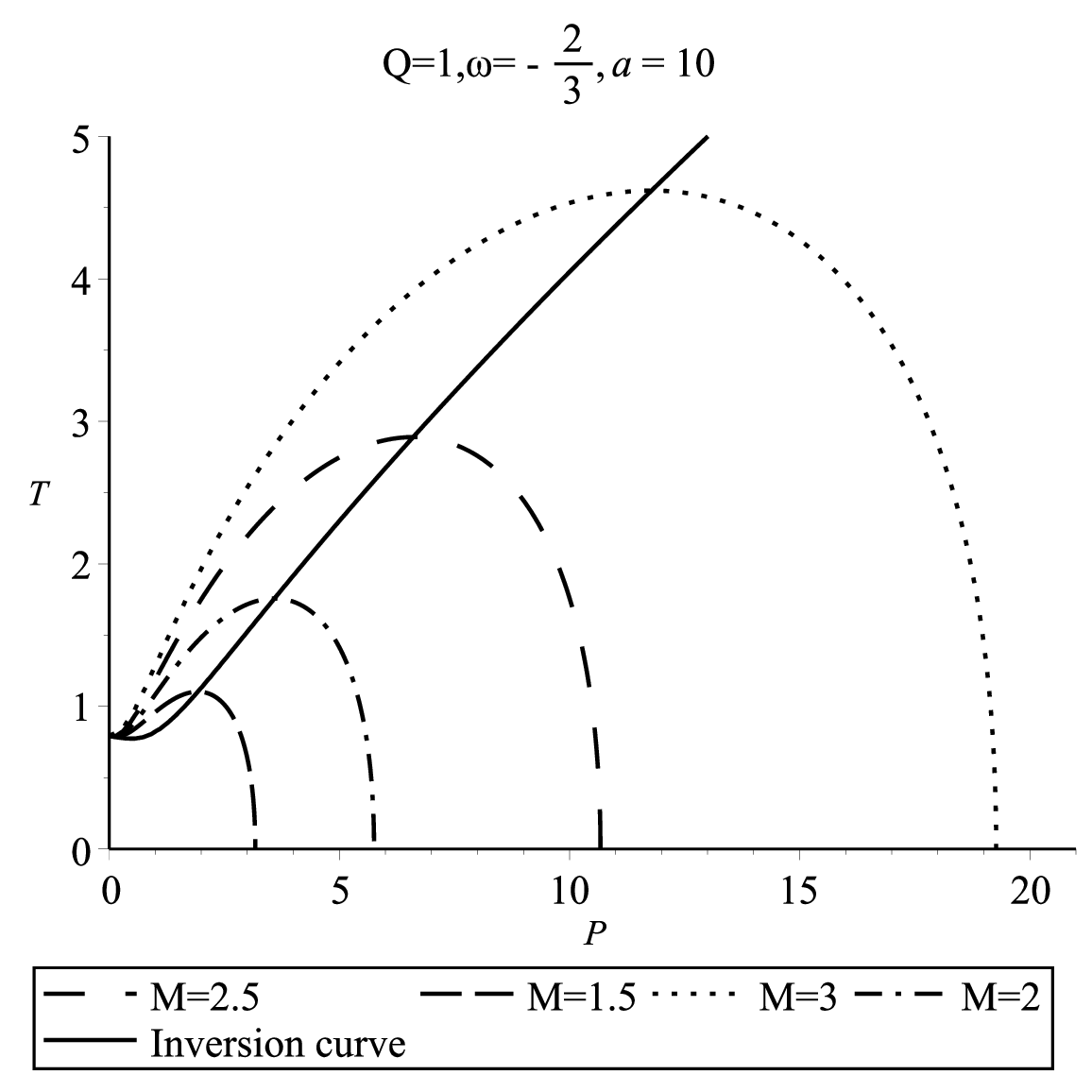}
 \includegraphics[width=6cm,height=4.5cm]{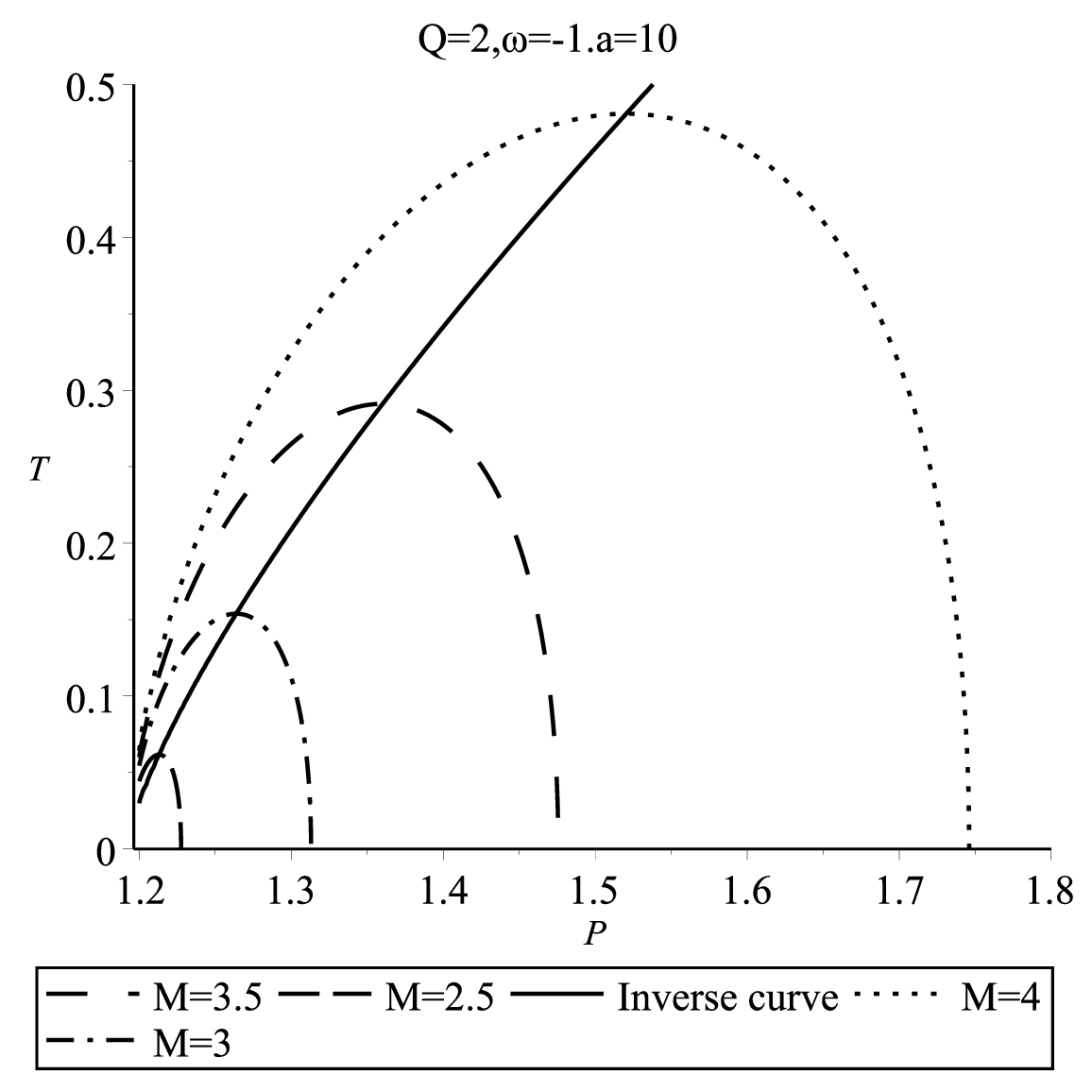}
 \includegraphics[width=6cm,height=4.5cm]{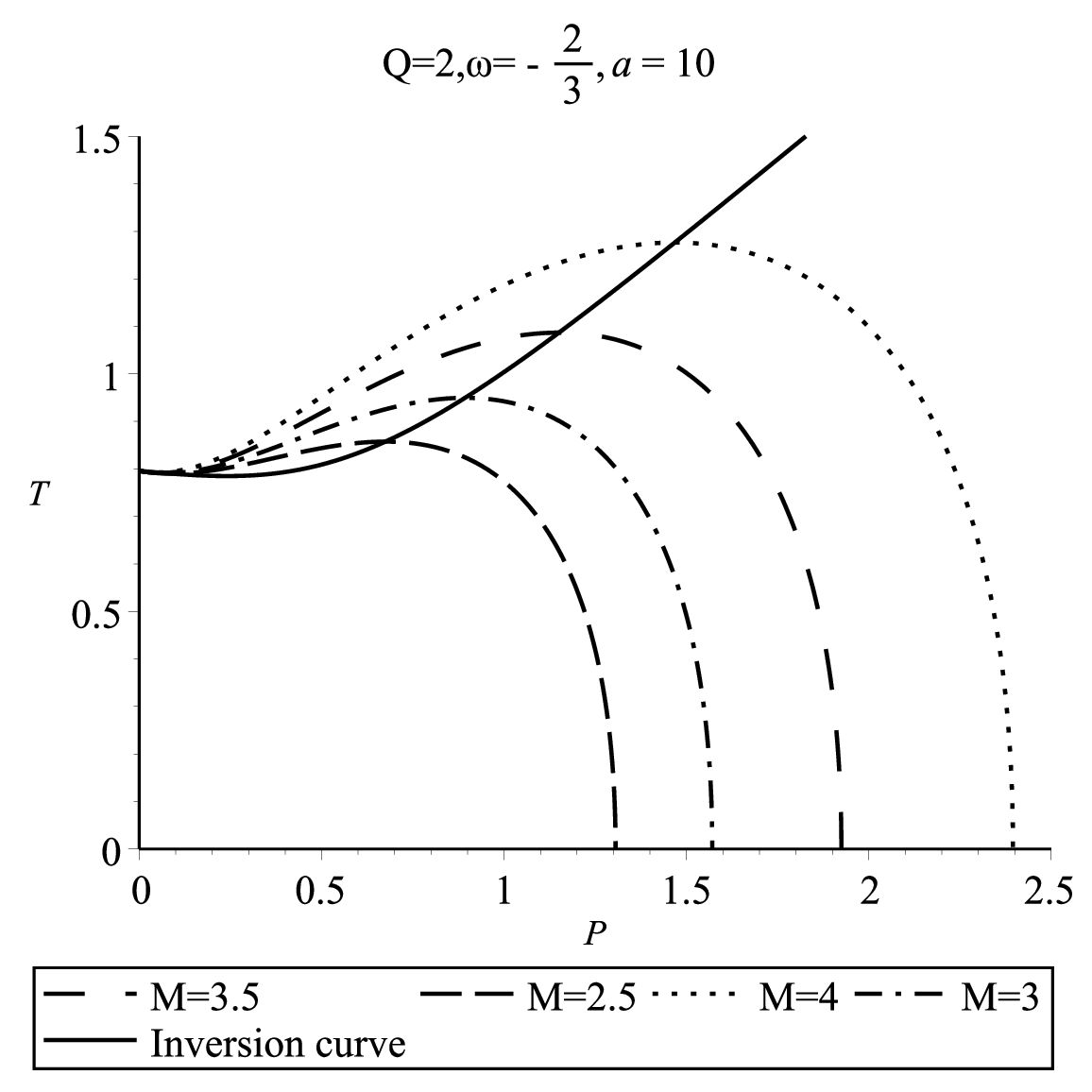}
 \includegraphics[width=6cm,height=4.5cm]{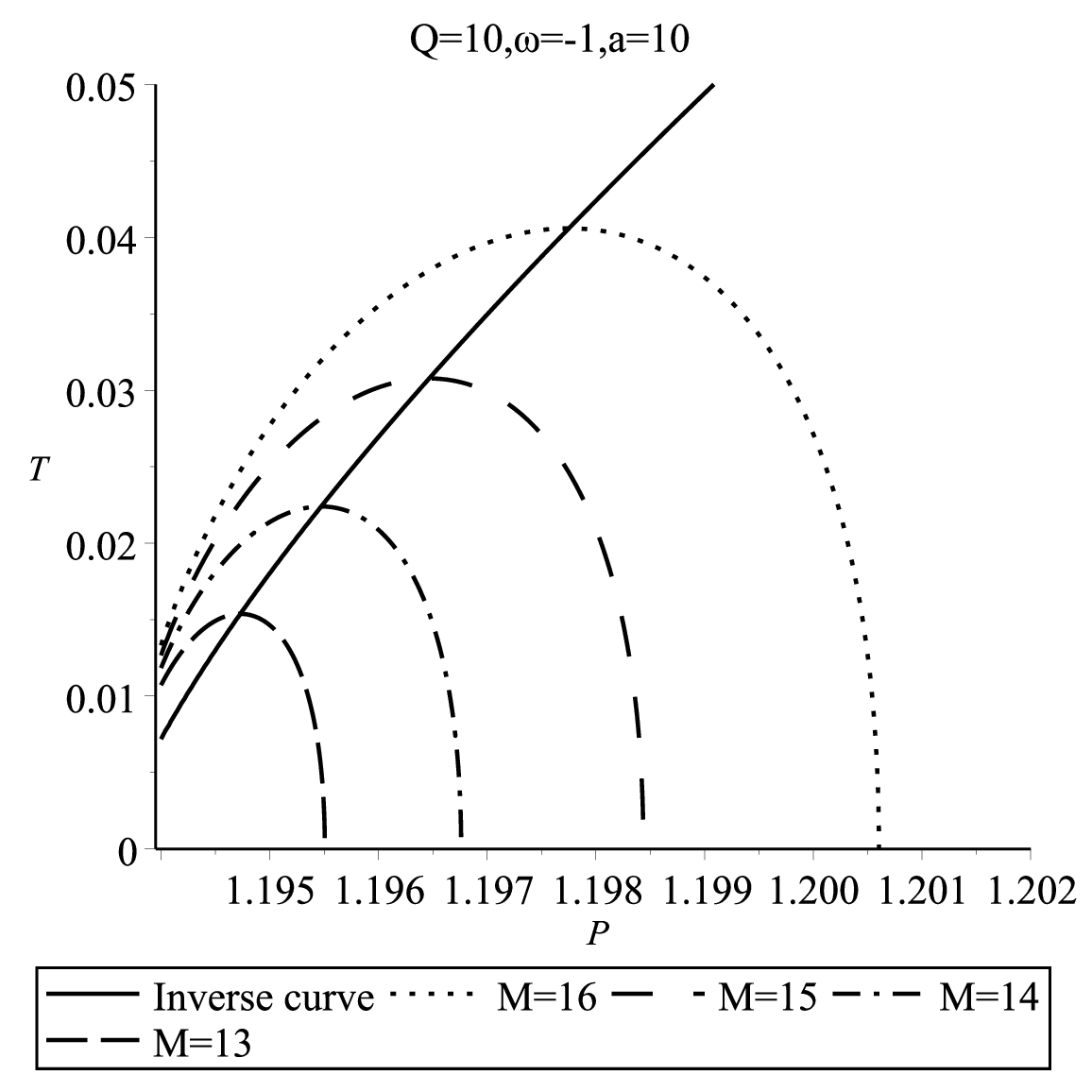}
\includegraphics[width=6cm,height=4.5cm]{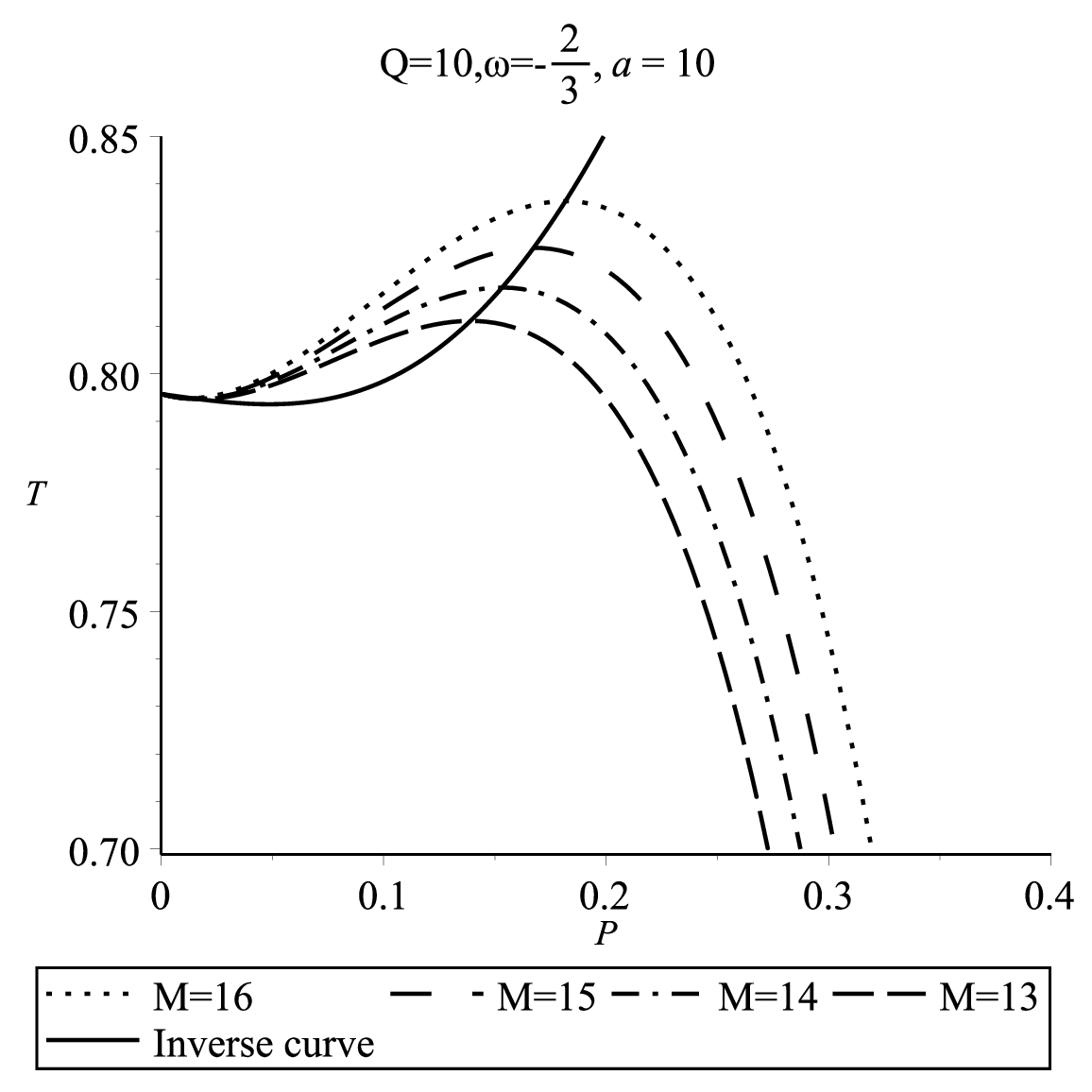}
 \includegraphics[width=6cm,height=4.5cm]{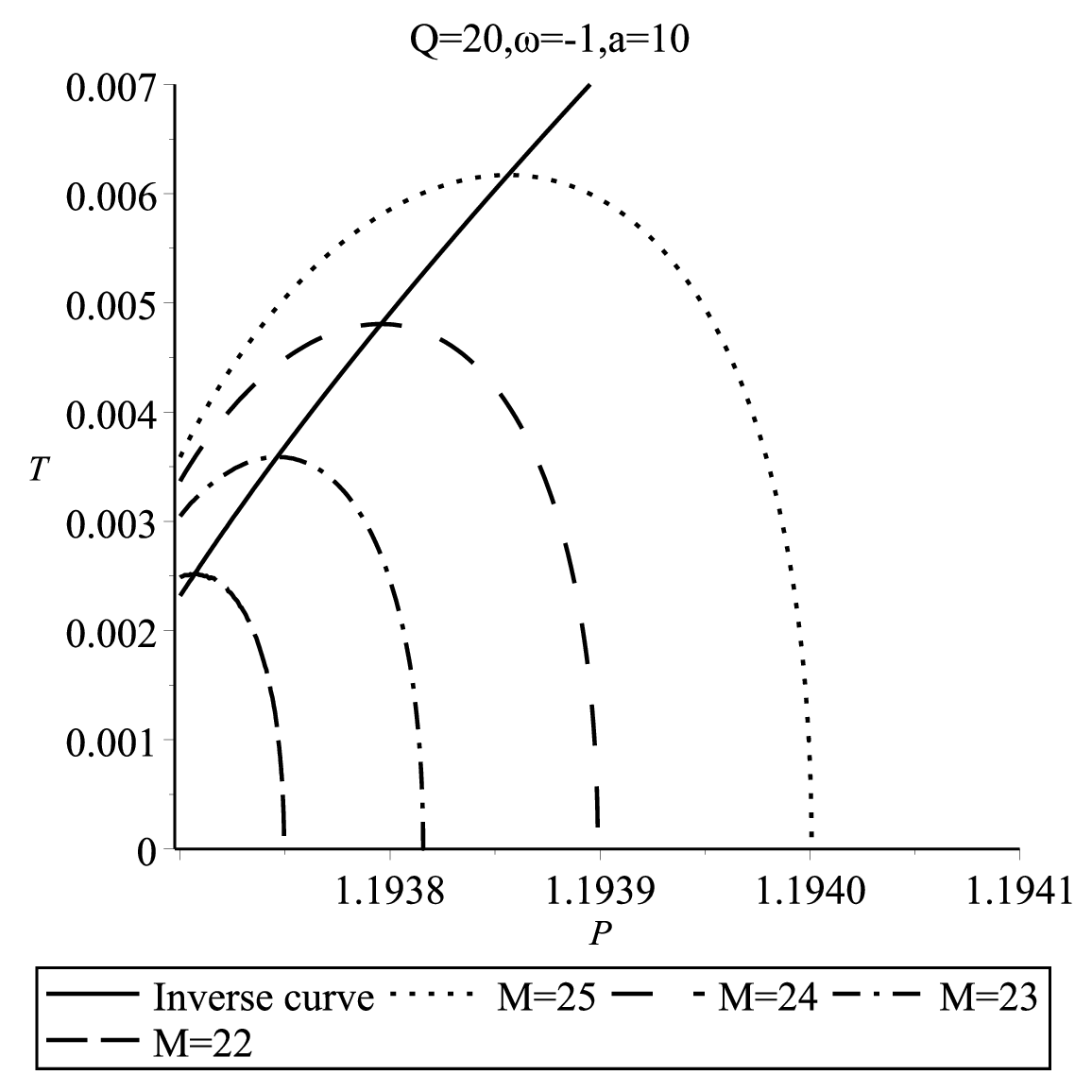}
\includegraphics[width=6cm,height=4.5cm]{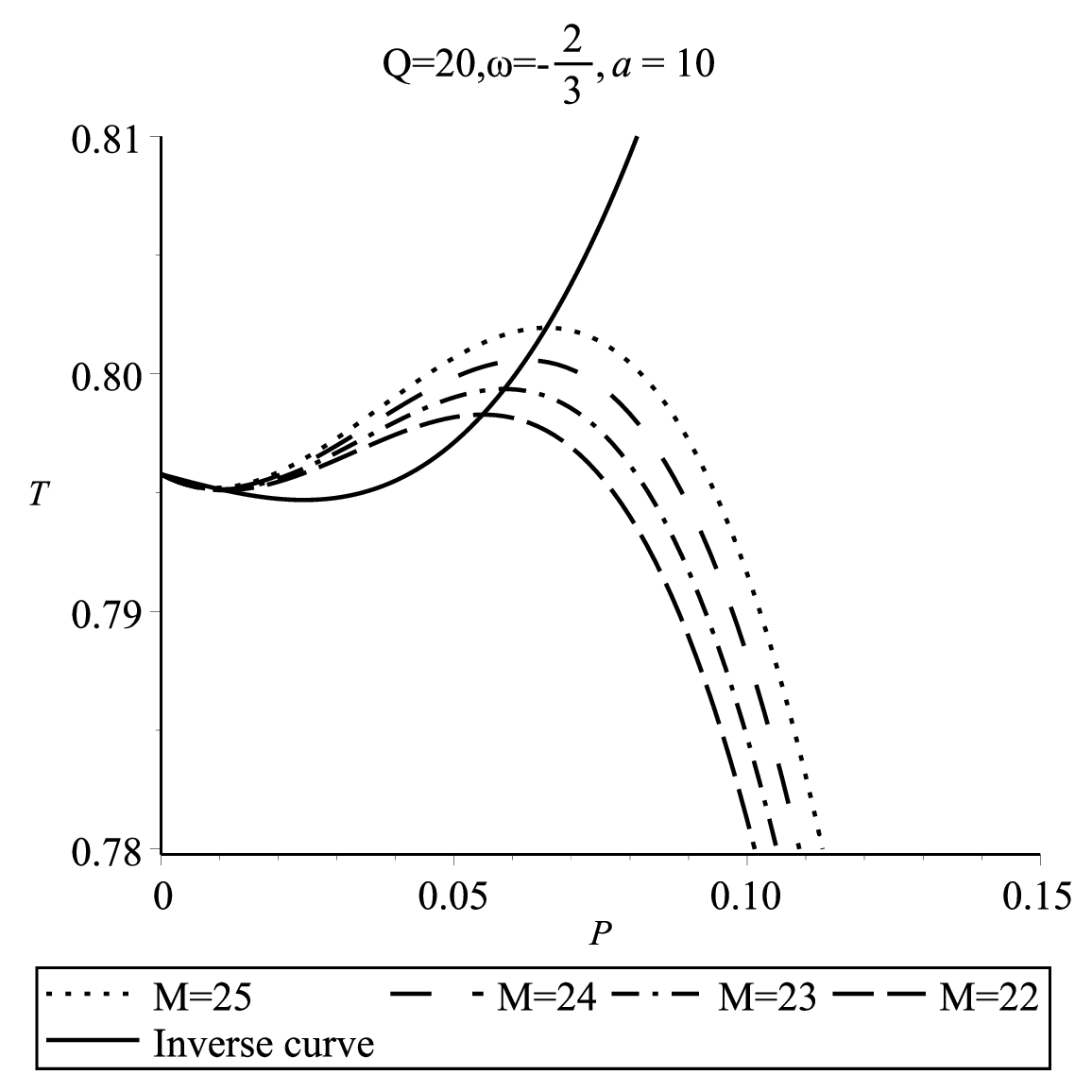}
\caption{\label{fig:1}   Diagrams of the inversion curves and the
isoenthalpic curves  of the  RNAdS black hole with the strong
effects of the quintessence $a=10.$}
\end{figure}

\end{document}